\documentclass[12pt]{iopart}
\usepackage{graphicx}
\usepackage{dcolumn}
\usepackage{bm}
\usepackage{xcolor}
\usepackage{subcaption}
\usepackage{float}
\usepackage{cite}
\usepackage[utf8]{inputenc}
\expandafter\let\csname equation*\endcsname\relax

\expandafter\let\csname endequation*\endcsname\relax
\usepackage{amsmath}
\usepackage{amssymb}

\begin{document}

\title[]{Super-resolution Optical Fluctuation Imaging---fundamental estimation theory perspective}

\author{Stanisław Kurdziałek$^{1}$ and Rafał Demkowicz-Dobrzański$^{1}$}
\address{$^{1}$ Faculty of Physics, University of Warsaw, Pasteura 5, PL-02-093 Warszawa, Poland}

\ead{s.kurdzialek@student.uw.edu.pl}

\vspace{10pt}

\begin{abstract}
We provide a quantitative analysis of super-resolution imaging techniques which exploit temporal fluctuations of luminosity of the sources in order to beat the Rayleigh limit. We define an operationally justified resolution gain figure of merit, that allows us to connect the estimation theory concepts with the ones typically used in the imaging community, and derive fundamental resolution limits that scale at most as the fourth-root of the mean luminosity of the sources.  We fine-tune and benchmark the performance of state-of-the-art methods, focusing on the cumulant-based image processing techniques (known under the common acronym SOFI), taking into account the impact of limited photon number and sampling time. 
\end{abstract}

%
%
%
%
%

\section{Introduction}
The wave nature of light imposes a limit on the resolution achievable by optical microscopes, 
known as the Rayleigh limit \cite{Born2013}. Nevertheless, over the past 40 years, many techniques, under the common name ``super-resolution imaging'' \cite{Betzig1995, Betzig2006, Rust2006, Hell2007, Dertinger2009, Bonnie2011, Monticone2014, Moerner2015,  Tsang2016, Paur2016, Yang2016, Tham2017, Parniak2018, Erkmen2008, Boto2000, Taylor2014, Rozema2014, Dowling2015, Genovese2016, Schnell2019}, have been developed to bypass this limit. Almost all far-field super-resolution methods can be divided into three groups depending on the way in which assumptions laying behind the derivation of the traditional resolution limits are broken: (i) sample (light emitters) modification \cite{Betzig1995, Betzig2006, Rust2006, Hell2007, Dertinger2009, Bonnie2011, Schwarz2012, Monticone2014, Moerner2015}, (ii) 
    outgoing light measurement modification \cite{Tsang2016, Paur2016, Yang2016, Tham2017, Parniak2018} or (iii) illuminating light modification with a particular focus on the use of non-classical states of light \cite{Erkmen2008, Boto2000, Taylor2014, Rozema2014, Dowling2015, Genovese2016}.

Methods (ii, iii)  were largely developed by theorists and their fundamental potential and limitations are well understood in terms of quantitative concepts from (quantum) information and estimation theories. In particular, by studying basic two (or few) point-sources imaging scenarios, optimal resolving protocols have been designed, and rigorous upper-bounds on achievable resolution gains derived.
Still, due to technical challenges, the practical impact of these methods is debated and the majority of experimental implementations are proof-of-principle demonstrations rather than versatile imaging systems. 

In contrast, methods (i)  have been largely developed by experimentalists, are commonly used in modern fluorescent microscopy, and are practical for imaging of 2D, or even 3D samples with an arbitrarily complex distribution of emitters. 
 A significant portion of these methods make use of temporal correlations of intensity of each emitter. In methods such as stochastic optical fluctuation imaging (SOFI) \cite{Dertinger2009}, stochastic optical reconstruction microscopy (STORM) \cite{Rust2006}, and photo activated localization microscopy (PALM) \cite{Betzig2006}, positive temporal correlations, explainable by a classical model of emitters with fluctuating brightness, are utilized. Negative, inherently quantum correlations  (anti-bunching) can be used to obtain super-resolution as well \cite{Schwarz2012, Monticone2014}. Despite their practical relevance, methods (i) have not been given as much estimation-theoretical attention  as methods (ii-iii), see \cite{Ram2006rohtua} for some notable exception.  The goal of this paper is to fill in this gap by providing a comprehensive study of the so called SOFI \cite{Dertinger2009}. The introduced framework can be, however, generalized to other super-resolution techniques, e.g. anti-bunching based method \cite{Schwarz2012, Monticone2014}, as we demonstrate in Section \ref{secantb}.

One of the main challenges in approaching imaging problems using the estimation theory perspective is the complexity of the imaging task when viewed as a multiple-parameter estimation problem \cite{Zhou2019, Albarelli2020}. As a result, an estimation based approach is usually restricted to rudimentary scenarios. The fundamental feature on which that estimation based studies focus on is the drop of  precision of estimation of the distance between two identical point sources as the distance is comparable or goes below the Rayleigh's limit. Furthermore, the variance of any unbiased estimator of sources separation tends to infinity as the separation goes to zero. It was already argued \cite{bettens1999model, Ram2006rohtua}, that such a statistical phenomenon, called Rayleigh's curse \cite{ Tsang2016}, is directly connected to the concept of the resolution. Recently, the method which in principle shows no sign of the Rayleigh's curse has been proposed \cite{Tsang2016} and implemented \cite{Parniak2018}. Unfortunately, Rayleigh's curse eventually always reappears for small enough separations in all realistic scenarios, when the presence of noise is assumed \cite{len2020resolution}. In practice, super-resolution techniques (including the one studied in this paper) don't provide non-vanishing estimation precision for zero separation, but they allow to enhance the precision for sub-Rayleigh distances. The similar effect can be achieved by modifying the imaging system, such that its Point Spread Function (PSF) gets narrowed. An important contribution of this letter is a proposal of the operationally meaningful quantity that relates the estimation precision enhancement achieved with a given super-resolution technique and the narrowing the PSF by an equivalent factor. This quantity, which will be defined further on in the text, allows to connect two different views on the resolution---the one based on the estimation theory, and the one related to the \textit{effective} PSF size. This will also allow us to interpret the resolution limits obtained when studying the two-point separation problem, as a valid (but possibly not tight) limits for imaging more realistic multiple point sources.

Thanks to this connection, it is possible to properly account for the effects of noise, among which the most fundamental is the shot noise resulting from the finite detection statistics.
The impact of shot noise is often far from obvious for more sophisticated algorithms of image reconstruction, and, as will be discussed below, cannot be ignored even when dealing with bright sources. Furthermore, when finite detection statistics is combined with the finite correlation times of fluctuating emitters, a non-trivial trade-off in the choice of the optimal sampling time arises---the longer time of a single frame, the better photon statistics, but at the same weaker inter-frame intensity fluctuations. 
\section{Estimation theory for optical imaging}
In this section we demonstrate how estimation theory tools can be used to provide a meaningful resolution gain figure of merit, which encompasses the effects of noise and the effective PSF size. We start with a brief review of the ideas laying behind the superresolving power of the SOFI technique to illustrate the traditional approach to quantifying super-resolution. 
\subsection{Basics of SOFI}

  The SOFI method is based on calculating temporal cumulants of measured intensity distribution in a number of time frames. It's often claimed, that the resolution can be increased by a factor $\sqrt{k}$ if the $k$-th cumulant is computed, which is justified as follows.
If the imaged sample consists of $L$ independently fluctuating point emitters, then the light intensity observed in the image plane is:
\begin{equation}
    I( \vec r, t) = \sum_{i=1}^L P_i(t) U(\vec r - \vec{r_i}),
\end{equation}
where the stochastic process $P_i(t)$ represents the fluctuating brightness of $i$-th emitter, $\vec{r_i}$ its position in the image plane, and $U(\vec r)$ is the PSF of the system. Since the emitters are independent, $k$-th temporal cumulant of the signal (at a given $\vec r$) reads:
\begin{equation}
    \kappa_k(\vec r) = \sum_{i=1}^L \kappa_k [P_i(t)] U^k(\vec r - \vec {r_i}),
\end{equation}
where $\kappa_k[P_i(t)]$ is the $k$-th cumulant of the stochastic process $P_i(t)$. 
The PSF is now replaced by its $k$-th power. If the standard, Gaussian approximation of the PSF is used, $U^k$ is narrowed by a factor $\sqrt k$ compared with $U$.

The described scheme can be improved in various ways. The most important modification is based on utilizing spatial correlations (cross-cumulants) in an image reconstruction algorithm \cite{Dertinger2010}. Cross-cumulants of $k$-th order are computed for all $k$-element subsets of all pixels, and the cumulant computed for a given set gives rise to the signal located in its centroid. Cross-cumulants corresponding to the same centroid can be summed directly, or with proper weights in order to maximize the signal-to-noise ratio (SNR) \cite{vandenberg2016model}. This approach not only allows to make use of information hidden in the spatial correlation, but also increases the number of pixels in the final image, which is significant from the practical point of view. The latter effect can be also achieved if recorded images are Fourier-interpolated before further processing \cite{stein2015fourier}.

Unfortunately, even after applying the described modifications, higher cumulants are more noisy, and it's not possible to achieve the unlimited resolution gain in practice---this effect reappears in all known experiments. Therefore, noise has to be taken into account in order to assess the maximal resolution gain achievable in SOFI. Some analysis of the impact of noise on the computed cumulants estimators have been made \cite{Wang2016, Vandenberg2019,   moeyaert2020sofievaluator}, but these studies have not employed estimation theory concepts such as the Fisher information (FI), and did not make an attempt to benchmark the performance of the methods against the fundamental limitations imposed by estimation theory. Our goal is to provide such a rigorous study.

\begin{figure}[t] 
\centering
\includegraphics[width=0.85 \columnwidth]{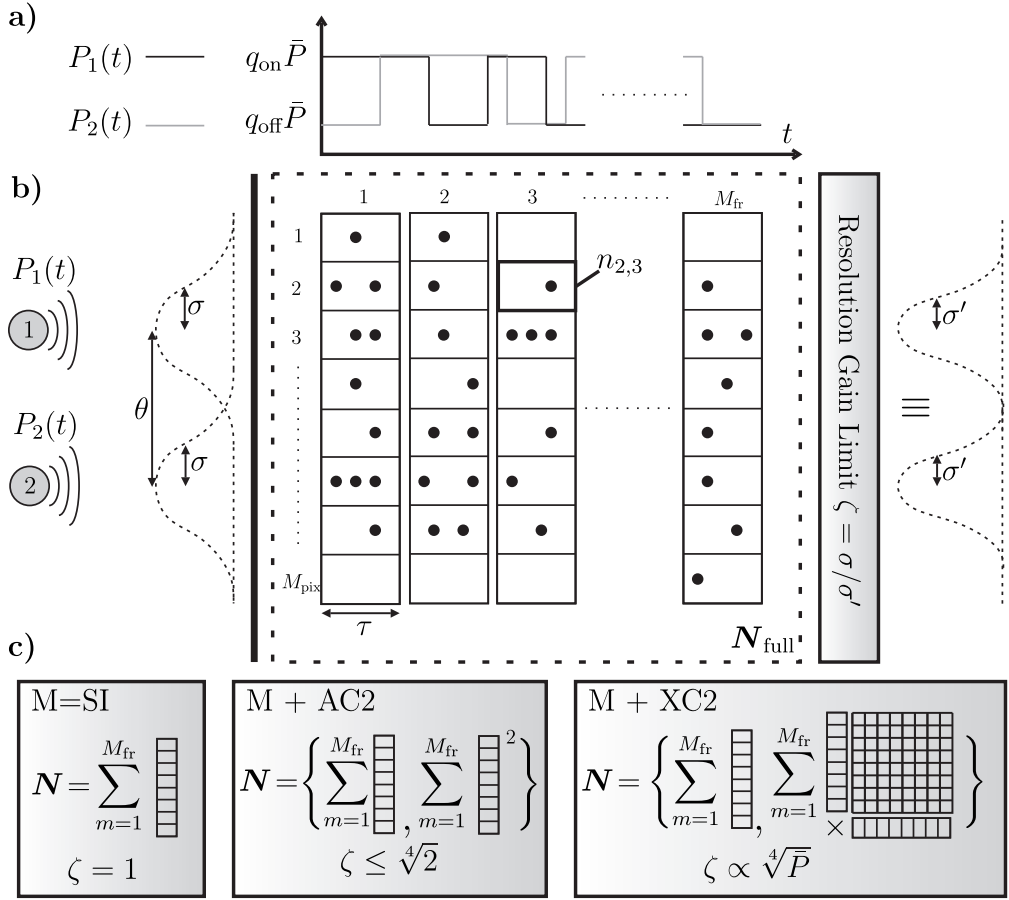}
\caption{Imaging model and the overview of the main results. a) Two point sources stochastically switching between two luminosity levels. b) Imaging task reduced to estimation of two point sources separation in the regime of overlapping point-spread-functions. By exploiting the full information $\boldsymbol{N}_{\textrm{full}}$ of number of photons $n_{i,m}$ registered in a given pixel and in a given time-frame of duration $\tau$, it is possible to provide an effective enhancement in resolution compared with the standard imaging where the numbers of photons measured in different time-frames are summed.
c) Conceptual representation of different 
reconstruction methods (utilizing incomplete data $\boldsymbol{N}$) and the corresponding resolution gain limits $\zeta$: M (mean intensity) = SI (standard imaging), M+AC2 (mean + second temporal auto-cumulant analysis), M + XC2 (mean + cross-cumulant analysis).
}
\label{fig:scheme}
\end{figure}
\subsection{Resolution Gain Limit}
Let's consider the simplest, yet representative case of imaging a binary object, which consists of two identical point emitters with fluctuating brightness. Those two emitters are assumed to lie on a known axis, perpendicular to the optical axis of the imaging system, so the  problem becomes 1D, and only transverse resolution is studied. Moreover, we assume that the centroid of the object is also known, and only the distance between emitters ($\theta$) needs to be estimated, see Figure ~\ref{fig:scheme} (note that in case of emitters of different brightness all the reasoning will be basically unaltered provided one replaces the geometric distance between the sources by a quantity based on the second moment of intensity distribution, as this is the quantity that is subject to the Rayleigh curse  \cite{Tsang2016, Chrostowski2017, Zhou2019}).  Given a random vector $\boldsymbol{N}$ that represents the data, distributed according to a probability distribution which is a function of the estimated parameter $p_\theta(\boldsymbol{N})$, the variance $ \textrm{Var} [ \tilde \theta]$ of any locally unbiased estimator of $\theta$ is 
lower bounded by 
$(\mathcal F_\textrm{meas})^{-1}$, where 
\begin{equation}
\label{Fisher}
    \mathcal F_\textrm{meas}(\theta) = \int  \frac{1}{p_\theta(\boldsymbol{N})} \left(\frac{  \partial 
    p_\theta(\boldsymbol{N})}{\partial \theta } \right)^2  \, \textrm{d} \boldsymbol{N}
\end{equation}
is the FI associated with the whole measurement \cite{Kay1993}. For the purpose of comparing different strategies we will use the FI per photon $\mathcal{F}(\theta) = \mathcal{F}_\textrm{meas}(\theta)/\bar{N}$, where $\bar{N}$ is the mean number of photons involved in the experiment. 

\begin{figure}[t]
\centering
\includegraphics[width=0.735 \columnwidth]{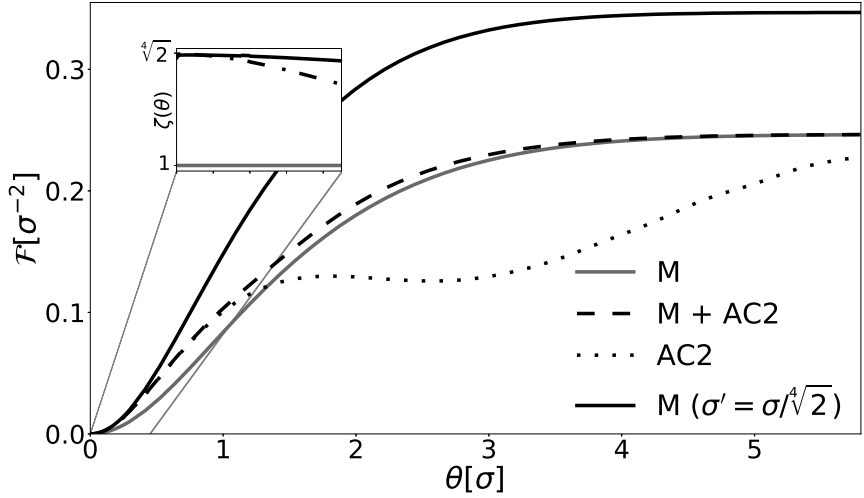}
\caption{Comparison of the FI associated with the mean intensity analysis (M), the FI for the estimation scheme involving the mean and the 2nd temporal cumulant (M+AC2) and the FI based on the 2nd auto-cumulant only (AC2). For small distance $\theta$, very bright emitters, and strong brightness fluctuations, improvements are equivalent to narrowing of the PSF by $\zeta = \sqrt[4]{2}$. Pixel size $\Delta x = 0.5 \sigma$ is assumed.}
\label{w1}
\end{figure}

From now on, we assume that the PSF is Gaussian with standard deviation $\sigma$. The FI per one photon for standard imaging (SI) of Poissonian sources with constant brightness \cite{Ram2006rohtua, Tsang2016},  $\mathcal F^\textrm{(SI)}$, is sketched as a function of $\theta$  in Figure ~\ref{w1}. A significant drop in estimation precision below the Rayleigh limit is visible. 
In super-resolution microscopy we are mostly interested in the sub-Rayleigh regime, i.e. we assume that $\theta \ll \sigma$. If no noise apart from shot noise is present, and the effect of finite spatial resolution of the detector is neglected, $\mathcal F^\textrm{(SI)}$ for small $\theta$ can be approximated as (see \ref{secA} for details) 
\begin{equation}
\label{FSI}
    \mathcal F^\textrm{(SI)}(\theta) =     \mathcal F^\textrm{(M)}(\theta)  = \theta^2 /8\sigma^4 + \mathcal O(\theta^4/\sigma^6),
\end{equation}
where we have also indicated that SI is equivalent to the analysis based only on the mean  of the total number of photons (M) collected over the whole duration of the experiment. 
Now, if the PSF is narrowed by a factor $s$, the FI in the limit $\theta \rightarrow 0$ increases by a factor $s^4$---this observation allows us to connect PSF-size approach with an estimation theory approach. If a given super-resolution imaging scheme leads to an increase of FI from $\mathcal F^\textrm{(SI)}(\theta)$ to  $\mathcal F(\theta)$, then for $\theta \ll \sigma$ this change is equivalent to narrowing of the PSF by a factor

\begin{equation}
\label{MRG}
    \zeta = \lim_{\theta \rightarrow 0}
    \left(\mathcal{F}(\theta) /\mathcal{F}^\textrm{(SI)}(\theta) \right)^{1/4}.
\end{equation}
The factor $\zeta$ will be called the Resolution Gain Limit (RGL) and will serve us as a figure of merit to asses the performance of different super-resolution methods.  

Admittedly, it does not contain all the information
on the performance of a given super-resolving technique, as it ignores 
the behaviour for larger $\theta$ (see Figure ~\ref{w1}), and the behaviour of the technique for more complicated objects. Nevertheless, this quantity captures in a simple way the essence of the super-resolving potential in a basic two point sources model, and allows to compare different methods in a well-defined way. Moreover, this quantity also provides us with a meaningful upper bound on the performance of a method in more complex imaging scenarios, as discrimination of binary objects is a prerequisite for resolving multiple sources. 
If a given super-resolving method yields a multiple-source image that  can be equivalently regarded as the one obtained with a standard imaging system with an appropriately narrowed PSF, then clearly when focusing on 
any two neighbouring points we can bound the performance of this method by our limit based on the simplified two-point imaging model. This limit may not be tight, but 
interestingly, even such an optimistic limit turns out to be lower than the resolution gain predicted by the naive PSF size analysis in some cases.
It is worth pointing out, that in all cases examined in this work, the ratio $\left(\mathcal{F}(\theta) /\mathcal{F}^\textrm{(SI)}(\theta) \right)^{1/4}$ decreases with increasing $\theta$. This implies that $\zeta$ 
remains a valid bound on the resolution gain enhancement irrespectively of the imaged points separation, and hence can be interpreted as a factor that reveals the maximal potential enhancement of a super-resolving method considered.
Finally, maximization of $\zeta$ 
in a given protocol can be regarded as a rule of thumb prescription on the choice of parameters that 
is likely to lead to the optimal performance of the protocol in real-life scenarios. 

\section{The Resolution Gain Limit for 2nd order SOFI}
\label{secsofi2}

In this section, we will introduce a realistic model of the emitters used in typical SOFI experiments. The results based on this model will be limited, due to its complexity, to image reconstruction methods based solely on 1st and 2nd order intensity correlations. The simplified model, which allows to extend the reasoning for higher-order intensity correlations, will be introduced in the next section.

For the binary source considered, we fix the positions of the emitters to be $-\theta/2$ and $ \theta/2$. Both emitters are statistically identical and independent. Fluctuations of a single emitter brightness are described by a stationary Markov process with two possible relative brightness levels $q_\textrm{on}$ and $q_\textrm{off}$ satisfying $q_\textrm{on}+q_\textrm{off}=1$, and $0 \le q_\textrm{off} \le q_\textrm{on} $. Such a description leads to exponential distributions for the occupation time of two states, which is observed for many typical dyes \cite{dickson1997off}, and can be used to approximate the QDs power-law blinking \cite{efros2016origin}. Two states have lifetimes equal to $\tau_\textrm{on}$ and $\tau_\textrm{off}$ respectively---in the examples studied we will set $\tau_{\textrm{on}} = \tau_{\textrm{off}} = \tau_0$ (some results for $\tau_\textrm{on} \ne \tau_\textrm{off}$ are shown in Section \ref{secasym}) and $\tau_0$ will play the role of an effective unit of time. The number of photons emitted from a single source over a short time $\delta t$, for which the relative brightness $q_i$ may be assumed to be fixed, is described by a Poisson distribution with mean $\bar P q_i \delta t$, where $\bar P$ parameterizes (in units $\tau_0^{-1}$) the average emitter brightness. Light is detected using a camera with a pixel size $\Delta x$ with the total number of pixels $M_{\textrm{pix}}$. No noise apart from shot noise is considered (additional background noise is analysed in Section \ref{secnoise}). 
In the analysed method it's crucial to track the time dependence of the light intensity, so the whole detection time is divided into $M_\textrm{fr}$ intervals of duration $\tau$, hereinafter called frames. In the end, one obtains a number of photons in each pixel and in each time frame $n_{i,m}$, where $i \in \{1,...,M_\textrm{pix}\}$ and $m \in \{1,...,M_\textrm{fr}\}$ stand for the pixel and the frame label respectively. In principle, $\theta$ may now be estimated from raw data $\boldsymbol{N}_{\textrm{full}}$ containing all $n_{i,m}$. At this point, however, we would like to consider scenarios in which particular algorithms of data analysis are used. We therefore construct a random vector $\boldsymbol{N}$ which contains combinations of variables $n_{i,m}$ which are  used in a given $\theta$ estimation procedure. Given the probability distribution family $p_\theta(\bm N)$, $\mathcal F$ can be computed using \eqref{Fisher}.

Let's restrict our considerations to vectors $\bm N$ of the form:
\begin{equation}
\label{6}
    \bm N = \frac{1}{M_\textrm{fr}} \sum_{m=1}^{M_\textrm{fr}} \bm{v}_m,
\end{equation}
where $\bm{v}_m=\left[v_{1,m}, \cdots, v_{n,m} \right]^T$ depends on variables $n_{1,m},n_{2,m},...,n_{M_{pix},m}$ only, in the same way for each frame. The simplest possible choice, $\bm{v}_m = \left[n_{1,m}, \cdots, n_{M_\textrm{pix},m} \right]^T$, corresponds to the SI approach, in which only the mean (M) value of signal is taken into account. 
To take advantage of fluctuations it's necessary to extend $\bm N$. If we choose $\bm{v}_m$, which consists of elements $n_{i,m}^k$ for $i \in \left\{1,...,M_\textrm{pix} \right\}$, $k \in \left\{ 1,2,...K \right\}$, it's possible to construct estimators based on the first $K$ 
auto-cumulants of the signal in each pixel (M+AC2+\dots+ACK), as well as compute the associated FI. This formalism also allows us to compute $\mathcal F$ when we restrict ourselves to the use of 2nd auto-cumulant only (AC2), as in the basic SOFI scheme (see \ref{secC} for details).

It's known, that the quality of the image in SOFI can be improved if the correlations between different pixels are utilized. In order to study the efficiency of these class of strategies, 
consider vector $\bm v_m$ comprising elements $ \left\{ \{ n_{i,m} \}, \{ n_{i,m} n_{j,m}  \} \right\}$ for $i,j \in \{1,2,...,M_\textrm{pix} \} $, which allows to compute a covariance estimator for each pixel pair (M + XC2). Note, however, that in a commonly used cross-cumulant based approach of image reconstruction in SOFI, one doesn't use each covariance independently. Instead, the covariances of pairs with the same centroid are summed, and such a sum is treated as a signal located at the given centroid \cite{Dertinger2010} (M+XC2s). In order to investigate, how much information is lost in such a summation, we will also compute $\mathcal{F}$ corresponding to 
$\bm v_m = \left[ S_{1,m}, S_{3/2,m}, \cdots, S_{M_\textrm{pix},m},
n_{1,m}, \cdots, n_{M_\textrm{pix},m}  \right]^T$, where $S_{l,m} = \sum_{(i+j)/2=l} n_{i,m} n_{j,m}$. 
\newline \newline
Note, that the elements of $\bm N$ are in general correlated in a very non-trivial way. However, things simplify in the limit $M_\textrm{fr} \rightarrow \infty$, as we can use the extended version of the central limit theorem \cite{Rosenblatt43} (valid in our case, when temporal correlations decay exponentially in time) to conclude that $\bm N$ is normally distributed. Consequently, the FI per photon
can be computed using the formula involving the mean value $\bm \mu$ of the distribution and its  covariance matrix $\bm \Sigma$ only \cite{Kay1993} 
\begin{equation}
    \mathcal F = \frac{1}{\bar{N}}\frac{\partial \bm \mu^\top}{ \partial \theta}  \bm \Sigma^{-1} \frac{\partial \bm \mu}{ \partial \theta},
\end{equation}
which is valid for $M_\textrm{fr} \rightarrow \infty$. $\bm{\mu}$ and $\bm{\Sigma}$ are computed with the help of the following formulas:
\begin{equation}
    \bm \mu = \left[  \left<v_{1,1} \right>, \left<v_{2,1} \right>, ...,\left< v_{n,1} \right>\right]^T,
\end{equation}
\begin{equation}
\bm \Sigma_{ij} = \frac{1}{M_\textrm{fr}} \left(  \textrm{cov}(v_{i,1},v_{j,1})+2 \sum_{m=2}^{\infty} \textrm{cov}(v_{i,1}, v_{j,m}) \right),
\end{equation}
where the homogeneity of the Markov processes was used. Notice, that correlations between frames affect $\bm \Sigma$, and therefore have impact on $\zeta$, even though statistics associated with these correlations are not directly used in the estimation scheme. See \ref{secC} for more details of $\bm \Sigma$ and $\bm \mu $ computation.

\begin{figure}[t]
\centering
\begin{subfigure}{.47\columnwidth}
  \centering
  \includegraphics[width=\linewidth]{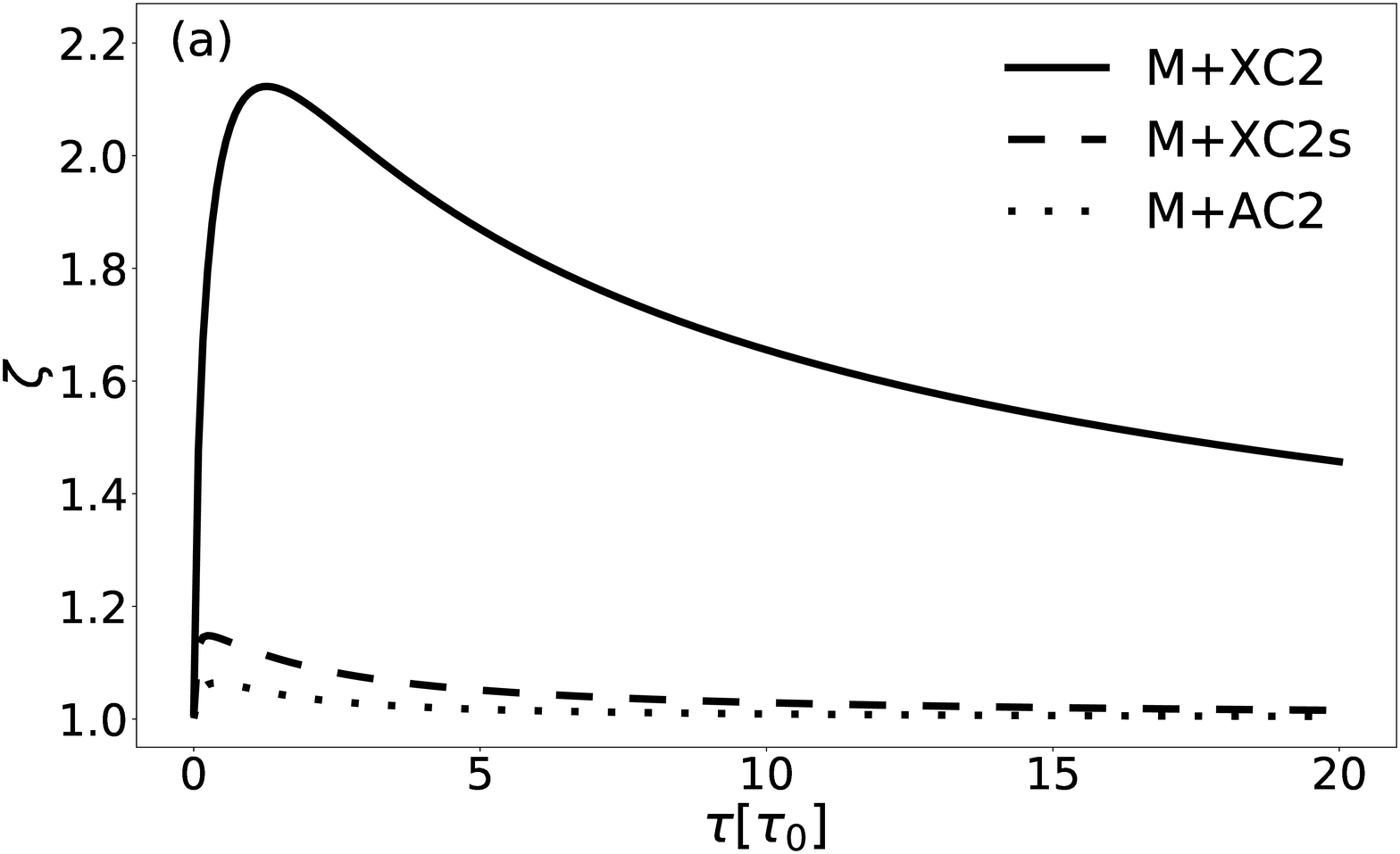}
  \label{w2_1}
\end{subfigure}
\begin{subfigure}{.47\columnwidth}
  \centering
  \includegraphics[width=\linewidth]{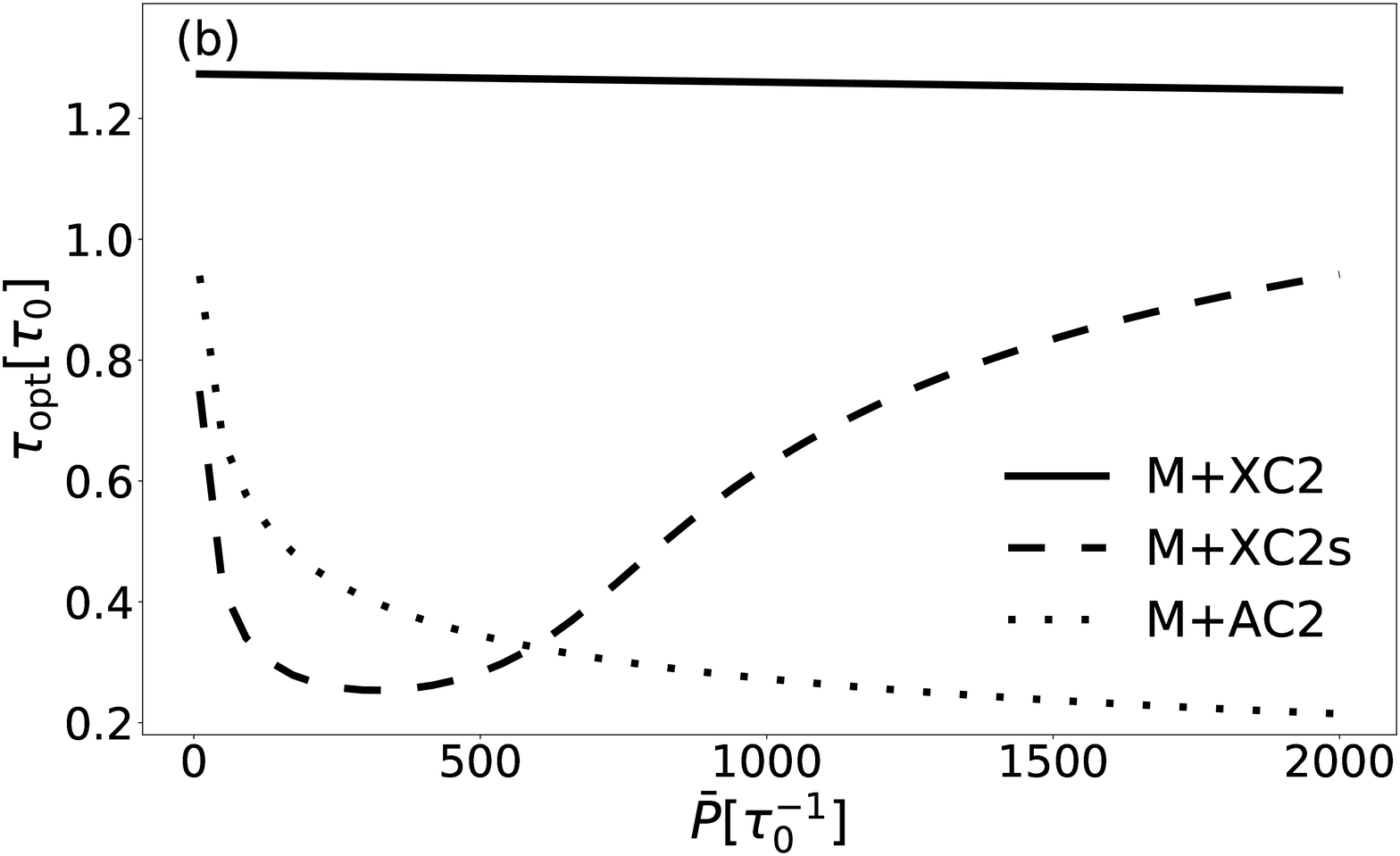}
  \label{w2_2}
\end{subfigure}
\caption{(a) Dependence of RGLs ($\zeta$) on the frame time $\tau$ (for $\bar P=300 \tau_0^{-1}$). (b) Dependence of the optimal time frame $\tau_\textrm{opt}$, for which $\zeta$ is maximal, on emitters brightness $\bar P$. Parameters used: $\tau_\textrm{on}=\tau_\textrm{off}=\tau_0$, $q_\textrm{off}=0$, $q_\textrm{on}=1$, $\Delta x = 0.5 \sigma$. }
\label{w2}
\end{figure}

We are now ready to compute $\mathcal F (\theta)$ for different estimation schemes, and check how the RGL defined in (\ref{MRG}) depends on the parameters of the setup. The way in which the RGL depends on the time of a single frame $\tau$ is particularly interesting. If $\tau$ is very long ($\tau \gg \tau_\textrm{on}, \tau_\textrm{off}$), then the fluctuations become averaged inside each frame, and can be hardly observed. On the other hand, when $\tau$ is too short, information contained in correlations between subsequent frames is lost. In the extreme case in which one photon is detected in a single frame at most, higher cumulants do not provide any extra information compared with the mean value of the signal. Detailed calculations confirm, that $\zeta \rightarrow 1$ in the limit $\tau \rightarrow 0$ and $\tau \rightarrow \infty$, both in the case of auto-cumulant and cross-cumulant based estimation. In order to reach the optimal $\zeta$ one needs to avoid both extremes and identify the optimal value of $\tau$, which in general depends on the estimation method  and the emitters brightness, see Figure ~\ref{w2}. 
Note that the cross-cumulant methods tend to benefit from longer frames, which allow to collect more photons and effectively reduce the shot noise of the data, while the auto-cumulant method, with its reduced data complexity, favours shorter frames and as a result stronger effective brightness fluctuations.  

\begin{figure}[t]
\centering
\begin{subfigure}{.47\columnwidth}
  \centering
  \includegraphics[width=\linewidth]{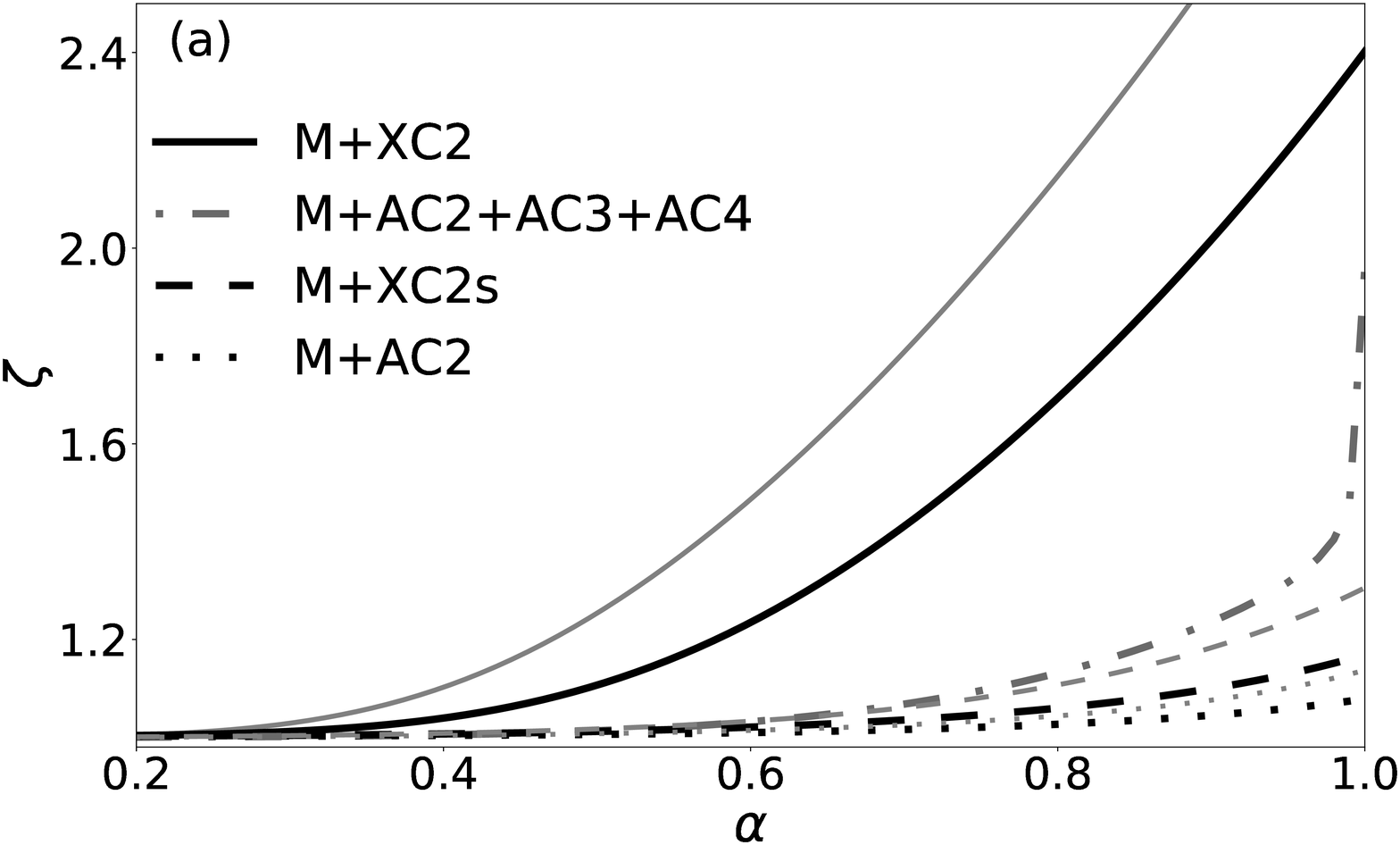}
  \label{w3_1}
\end{subfigure}%
\begin{subfigure}{.47\columnwidth}
  \centering
  \includegraphics[width=\linewidth]{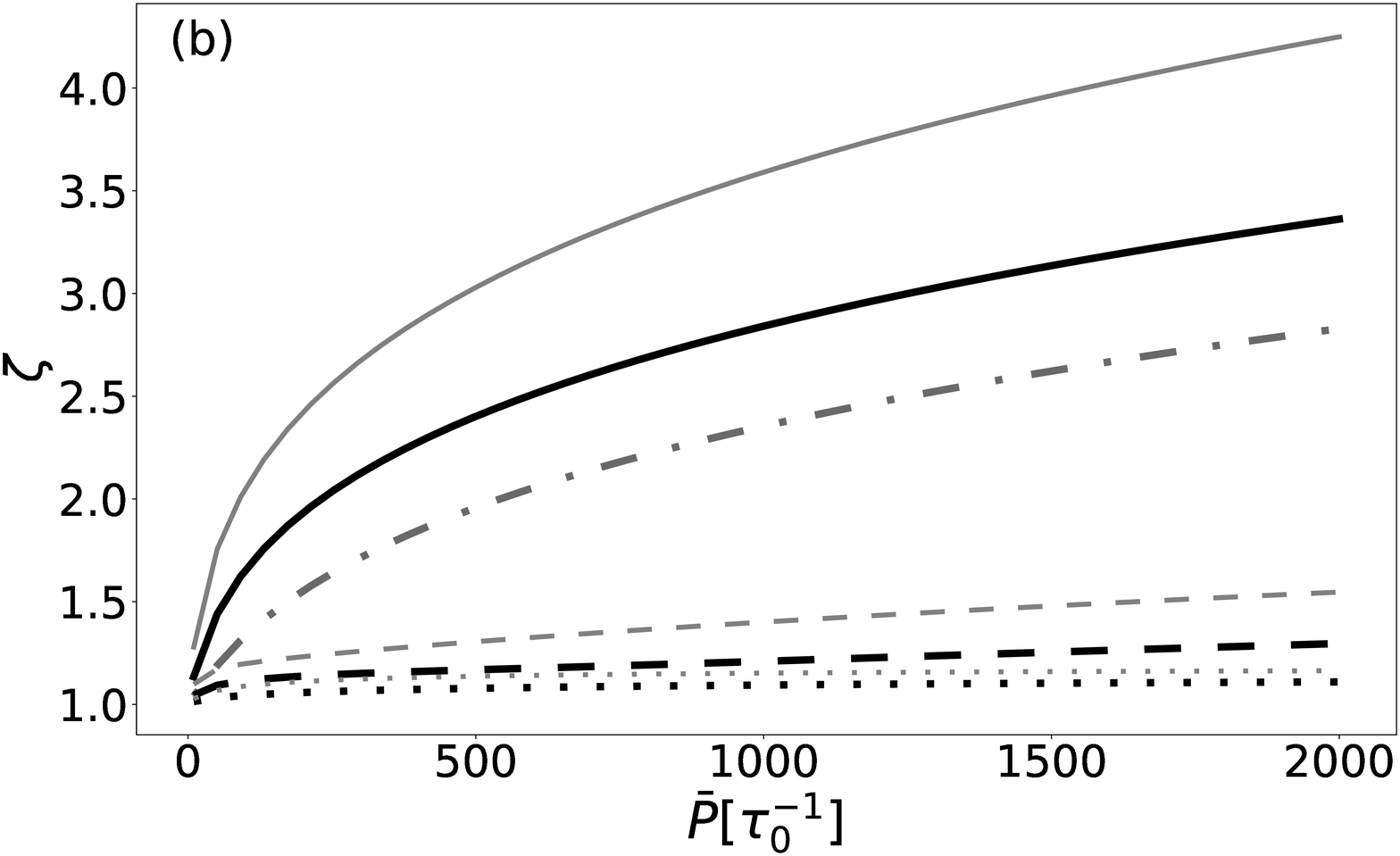}
  \label{w3_2}
\end{subfigure}
\caption{$\zeta$ as a function of $\alpha$ (for $\bar P =500\tau_0^{-1}$) (a), and as a function of $\bar P$ (for $\alpha = 1$)  (b), for different estimation schemes. Black lines correspond to results obtained using realistic blinking model, with $\tau_\textrm{on} = \tau_\textrm{off}=\tau_0$ and $\tau = \tau_\textrm{opt}$. Results obtained with the help of the simplified model ($p=0.5$, $\tau=\tau_0$) are denoted by gray lines.   Pixel size: $\Delta x = 0.5 \sigma$. }
\label{w3}
\end{figure}

Fixing the optimal frame time  $\tau = \tau_{\textrm{opt}}$, the dependence of RGLs on emitters brightness $\bar P$ and fluctuation strength defined as $\alpha = 1-q_\textrm{off}/q_\textrm{on}$ is shown in Figure ~\ref{w3}. 
The cross-correlation based approach outperforms the auto-cumulant based estimation significantly. Moreover, the relevant part of information is lost if the summation of covariances for pairs with the same centroid is carried out, as in \cite{Dertinger2010}. The use of a properly weighted sum, as proposed in \cite{vandenberg2016model}, allows to increase the RGL only slightly, as we demonstrate in Section \ref{sec3}. This indicates a space for improvement in the application of cross-cumulant based methods.

\section{Simplified fluctuations model and maximal RGL}
Until now, we have focused on estimation schemes based on 2nd order correlations. Going beyond this approach, we want to establish the fundamental upper-bound on the RGL, $\zeta_\textrm{max}$, which doesn't depend on the estimation scheme. To do so, we should compute the FI for a model involving all the data  $n_{i,m}$, and moreover, allow both the temporal and spatial resolution of the detector to be unlimited. This task is computationally much more challenging  than the previous one, so in what follows we consider a simplified model of fluctuating sources. 

Previously, the intervals between subsequent state switches were irregular. Therefore, brightness changes were observed within individual frames, and frames were correlated. From now on, we are going to neglect both of these effects, and assume, that brightness of both emitters are drawn in each frame independently, and remain constant in each frame taking values $q_\textrm{off}$, $q_\textrm{on}$ with probabilities $p$, $1-p$ respectively. The number of photons emitted during a single frame from a source with a relative brightness $q_i$ is drawn from the Poisson distribution with mean $q_i \bar P \tau$. 

Let us now check how the described simplification affects our previous results in the particular case in which the on- and off-states are equally probable. It corresponds to $p=0.5$ in the simplified model, and to $\tau_\textrm{on} = \tau_\textrm{off}$ in the  realistic one. We choose our parameters such that the average blinking frequency is the same in both models, i.e. the frame time in the simplified model is equal to emitters lifetimes in the Markov-process-based model ($\tau = \tau_\textrm{on} = \tau_\textrm{off}$). Frame time in the realistic model is assumed to be optimal $\tau = \tau_\textrm{opt}$. 
As can be seen in Figure ~\ref{w3}, the simplified model tends to overestimate $\zeta$, but qualitatively the dependence of $\zeta$ on different parameters as well as ordering of different methods in terms of their performance is unaffected.

\begin{figure}[t]
\centering
\begin{subfigure}{.47\columnwidth}
  \centering
  \includegraphics[width=\linewidth]{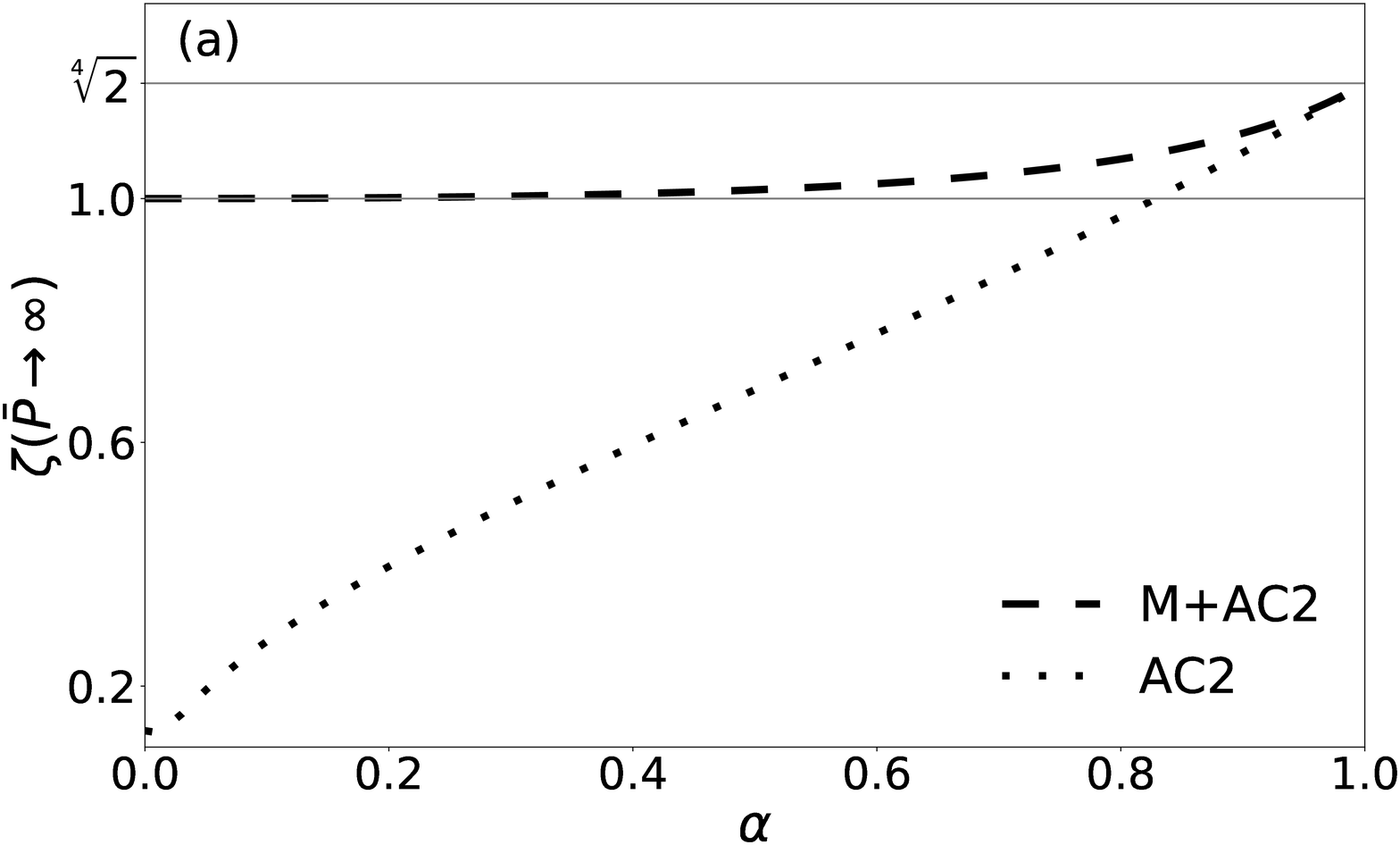}
  \label{w4_1}
\end{subfigure}%
\begin{subfigure}{.47\columnwidth}
  \centering
  \includegraphics[width=\linewidth]{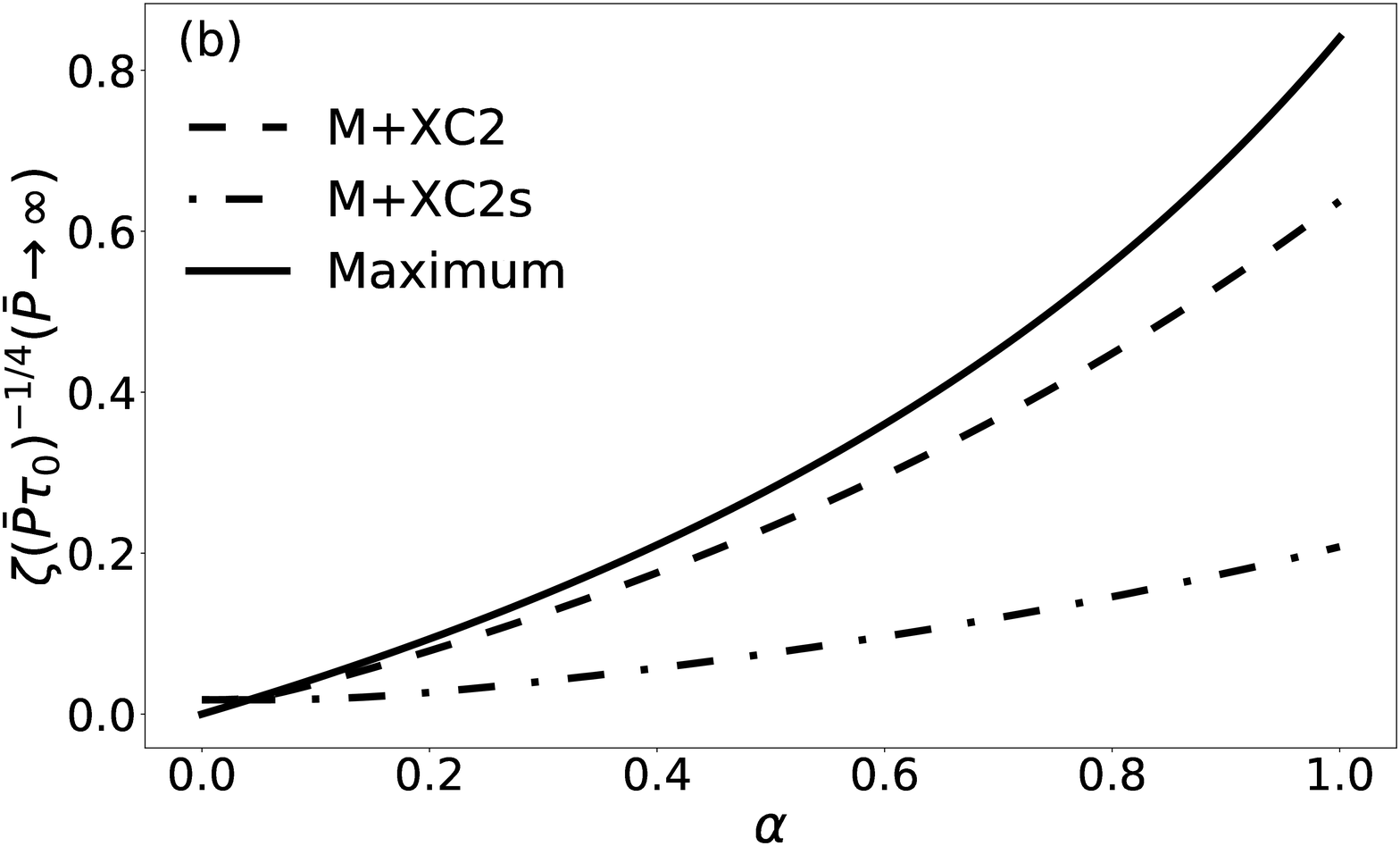}
  \label{w4_2}
\end{subfigure}
\caption{(a) $\zeta$ in the limit $\bar P \rightarrow \infty$ for the 2nd auto-cumulant based methods as a function of $\alpha$ for $p=1/2$ (calculated for the simplified model). 
RGL is never larger than $\sqrt[4]{2}$. 
Even for large $\bar P$, the replacement of mean with 2nd cumulant is not advantageous, unless fluctuations strength $\alpha$ is large enough ($\zeta <1$ for $\alpha < 0.83$). 
(b) Fundamental upper-bound $\zeta_\textrm{max}$ as well as RGLs associated with utilizing cross-cumulants, rescaled by $(P \tau_0)^{-1/4}$ to indicate the asymptotic behaviour when 
$\bar P \rightarrow \infty$.}
\label{w4}
\end{figure}
Unlimited spatial resolution of the detector means, that our complete data from each frame comprises the list of all the detected photon positions $x_1,...x_n$ (detection times do not provide any extra information in the model).  
Let's assume for a moment that relative brightness of two emitters ($q_1$,$q_2$) are fixed, and the total number of photons in a frame $n$ is known \textit{a priori}. Then the conditional probability of measuring a given photon positions sequence $x_1, ...,x_n$ is
\begin{equation}
\label{10}
    p_\theta(x_1,...,x_n|q_1,q_2,n) = \prod_{i=1}^n p_\theta(x_i|q_1, q_2),
\end{equation}
where
\begin{equation}
\label{11}
    p_\theta (x_i|q_1, q_2) = \frac{ q_1 U \left(x_i+\theta/2 \right) + q_2 U \left( x_i-\theta/2 \right)}{q_1 + q_2 }.
\end{equation}
The above formulas reflect the fact, that subsequent photons positions are uncorrelated if brightness are fixed, and the probability that a given detected photon was emitted from a given source is proportional to its brightness. In reality, one doesn't have a direct access to  relative brightness values $q_1, q_2$. The observed Probability Density Function (PDF) is averaged over unknown brightness:
\begin{equation}
\label{12}
    p_\theta(x_1,...,x_n|n) = \sum_{q_1, q_2 \in \{q_\textrm{on}, q_\textrm{off} \}} p_\theta(x_1,...x_n|n,q_1,q_2) P(q_1,q_2|n),
\end{equation}
where the conditional probability $P(q_1, q_2|n)$ is calculated using Bayes' formula:
\begin{equation}
\label{13}
    P(q_1, q_2|n) = \frac{P(n|q_1, q_2)P(q_1)P(q_2)}{P(n)}.
\end{equation}
Let's notice that photons positions drawn from the PDF (\ref{12}) \textit{are} correlated within a single frame---correlations arise when the information about brightness is hidden. The information about the total number of photons per frame $n$ is of course available, so it's possible to calculate the FI per frame for each fixed $n$ separately ($\mathcal F_{(n)}$), and then compute the FI per one photon using the formula
\begin{equation}
\label{mF1}
 \mathcal F = \frac{\left< \mathcal F_{(n)} \right>_n}{\left< n\right>_n},   
\end{equation}
where $\left< X(n) \right>_n \equiv \sum_n X(n) P(n)$ denotes averaging over $n$.
 The above procedure is used to obtain the expression for the complete-data-based FI (see \ref{secB} for details). The corresponding RGL reads 
\begin{equation}
\label{zetamax1}
    \zeta_\textrm{max} = \sqrt[4]{ 1 + G(p,\alpha, \bar P \tau) \bar P \tau },
\end{equation}
where $\alpha = 1-q_\textrm{off}/q_\textrm{on}$ is the fluctuation strength, and $G(p,\alpha,\bar P \tau)$ is explicitly defined in \eqref{gdef}.
For our analysis, it's crucial that function $G$ is ascending with respect to $\bar P \tau$ and is upper-bounded by an expression which does not depend on $\bar P \tau$
\begin{equation}
\label{zetamax2}
    \lim_{\bar P \tau \rightarrow \infty} G(p,\alpha, \bar P \tau) = \frac{\alpha^4 p (1-p)}{(2-\alpha)^3 (1-p \alpha)}.
\end{equation}
Numerical computations show, that the above limit approximates $G$ with an accuracy better than $1 \%$ for $\bar P \tau \gtrsim 2500$ in the case of weak fluctuations ($\alpha =0.2$), and 
$\bar P \tau \gtrsim 50$ for strong fluctuations ($\alpha=1$) (see Figure \ref{f4} ), in  which case the RGL can be approximated as:
\begin{equation}
\label{zetamax3}
    \zeta_\textrm{max} \simeq \sqrt[4]{1+p \bar P \tau}.
\end{equation}
From (\ref{zetamax1}) and (\ref{zetamax2}), we see that $\zeta_\textrm{max} \sim (\bar P \tau) ^{1/4}$ for large number of photons per frame $\bar P \tau$ for different fluctuation parameters. A similar scaling is observed numerically for $\zeta$ associated with cross-cumulant based estimation, see Figure ~\ref{w4}. If, however, only the mean and the 2nd auto-cumulant of the signal is involved in the estimation scheme, $\zeta(\bar P \tau = \infty) $ is finite. In particular, if we restrict our considerations to the symmetric case $p=0.5$ (other $p$ values are discussed in Section \ref{secasym}), no RGL higher than $\sqrt[4]{2}$ can be achieved. This demonstrates, that the resolution gain $\sqrt{2}$ predicted by the PSF narrowing analysis cannot be achieved even for strong fluctuations, very bright sources and large number of frames. The standard imaging scheme with a PSF narrowed by a factor larger than $\sqrt[4]{2}$ will always outperform 2nd auto-cumulant based SOFI if the basic task of resolving two point sources is considered. Interestingly, a similar discrepancy (the RGL is $\sqrt[4]{2}$, not $\sqrt{2}$) can be observed for a simple case of anti-bunching based imaging, as we will demonstrate in the next section.
\section{Practical aspects and extensions of the model}
\subsection{ The role of a pixel size}
\begin{figure}[ht]

\begin{subfigure}{.5\linewidth}
  \centering
  \includegraphics[width=\linewidth]{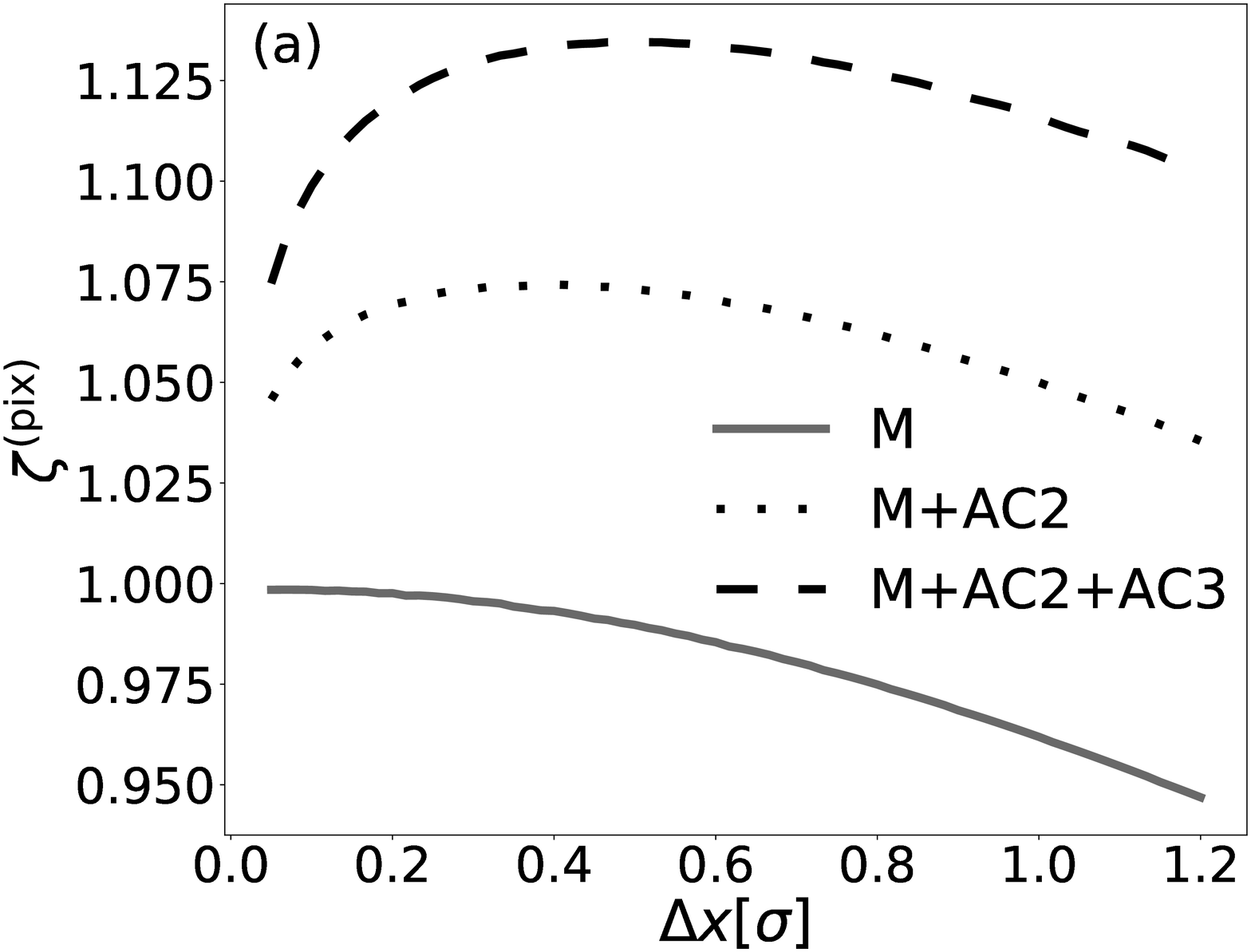}
  \label{f1}
\end{subfigure}%
\begin{subfigure}{.5\linewidth}
  \centering
  \includegraphics[width=\linewidth]{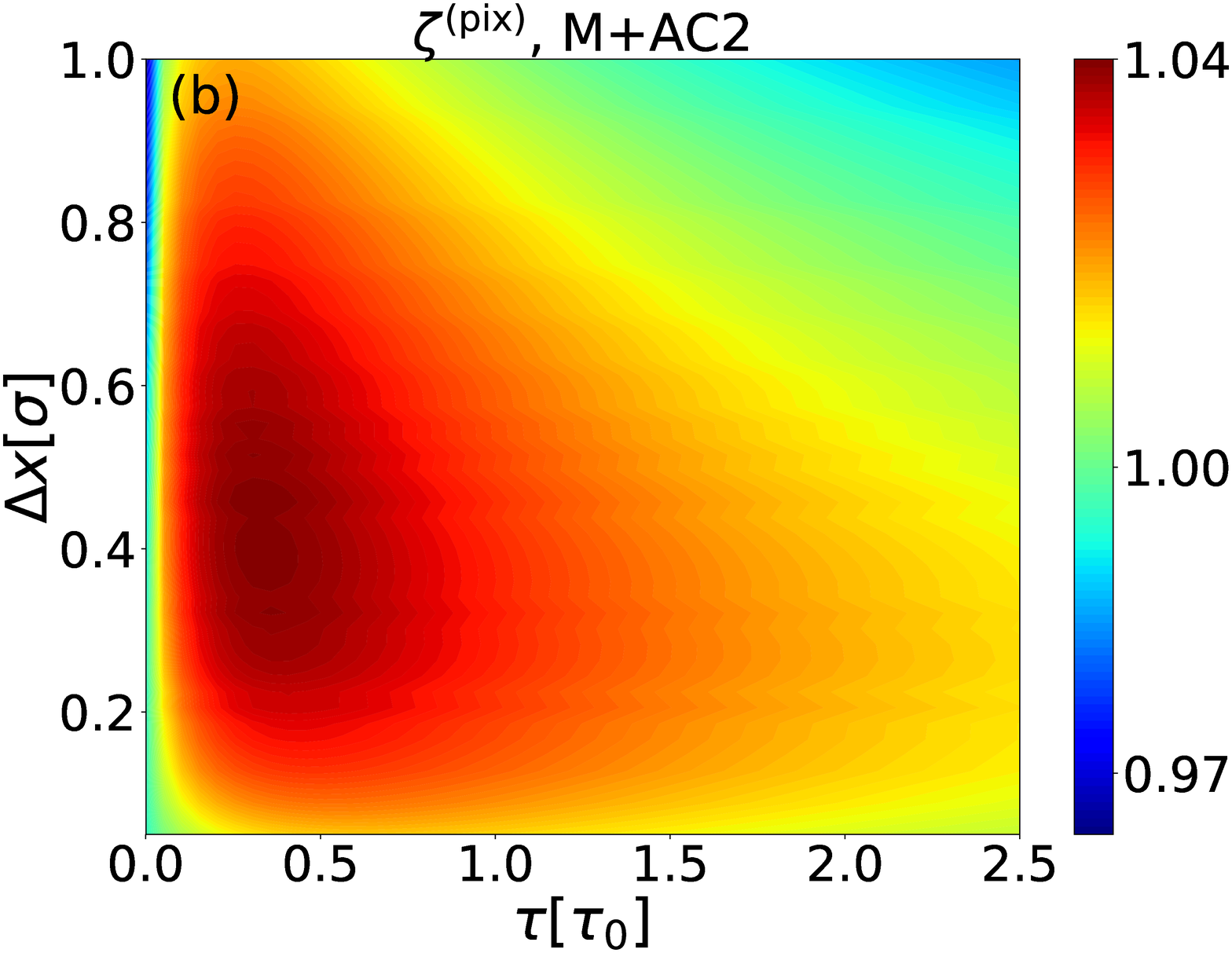}

\end{subfigure}

\caption{Figure (a) shows the dependence of $\zeta^\textrm{(pix)}$ on the pixel size $\Delta x$ for auto-cumulant based estimation schemes. If we go beyond the standard approach (M), infinitely small pixels are not optimal. Figure (b) shows how $\zeta^\textrm{(pix)}$ is affected when both $\Delta x$ and $\tau$ changes for 2nd auto-cumulant based estimation (M+AC2). In (a) the simplified model was used ($p=0.5$, $\bar P \tau =1000$), whereas for (b) Markov process based model with $\tau_\textrm{on} = \tau_\textrm{off} = \tau_0$, and $\bar P = 1000 \tau_0^{-1}$ was applied. In both cases $\alpha=0.9$. }
\label{s1}
\end{figure}

In the definition of $\zeta$, \eqref{MRG}, the same pixel size is used to calculate $\mathcal F$ and $\mathcal F^\textrm{(SI)}$. In order to examine the impact of the pixel size for different methods, we will use a slightly modified figure of merit defined as:
\begin{equation}
\zeta^\textrm{(pix)} = \lim_{\theta \rightarrow 0}
    \left(\mathcal{F}(\theta) /\left(\frac{\theta^2}{8 \sigma^4} \right) \right)^{1/4}.
\end{equation}
This modification fixes the denominator to $\mathcal F^\textrm{(SI)}$ associated with the infinite spatial resolution of the detector. It allows us to observe, for example, how the Standard Imaging resolution decreases when $\Delta x $ is too large. The role of the pixel size becomes less trivial when auto-cumulants are used in the estimation (see Figure  \ref{s1}). Pixels can't be of course too large, but very small pixels are no longer the optimal choice because the information contained in the correlations between pixels is lost. In particular, higher auto-cumulants don't provide any extra information if one photon per frame per pixel is detected at most. The described problem disappears when cross-cumulants are used, and very small pixels again become advantageous. Notice the fact, that a similar trade-off is observed in the time domain, when one changes the frame time. Very short time frames would only be optimal if correlations between frames were used, but such schemes are not analyzed in this work.

\subsection{Extra background noise}
\label{secnoise}
Even though the shot noise, resulting from the finite detection statistics, is the most fundamental one, other types of noise (e.g. camera noise) often play a significant role in SOFI.  Let us study a simple model, in which background noise is Poissonian, uncorrelated, and its mean value $\mu_B$ is the same in each pixel and in each frame. In order to take such a noise into account, it's enough to repeat the whole reasoning described in \ref{secsofi2} with only one modification---the term associated with the background noise should be added to each random variable $n_{i,m}$ (see \ref{secC} for more details).

\begin{figure}[h]
\centering
\includegraphics[width=0.735 \columnwidth]{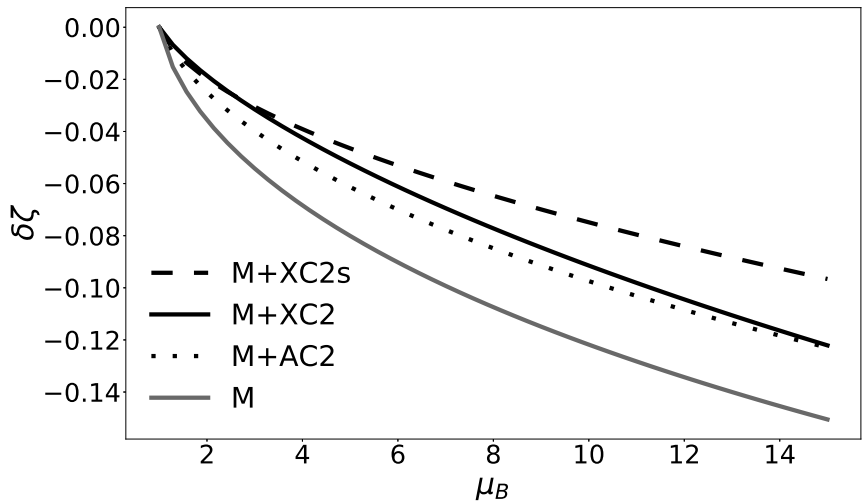}
\caption{The relative decrease of the RGL, $\delta \zeta = \frac{\zeta(\mu_B) - \zeta(\mu_B=0)}{\zeta(\mu_B=0)}$ is plotted as a function of $\mu_B$---mean background noise per pixel per frame. The simplified (independent frames) model is used, with parameters $\Delta x = 0.5 \sigma$, $\bar{n} = 1000$, $p=0.5$, $\alpha=1$. }
\label{s9}
\end{figure}

Non-zero value of $\mu_B$ leads to $\zeta$ decrease in all examined cases (where standard imaging without background noise is treated as a reference when $\zeta$ is computed). It turns out that cumulants based methods are more robust against this type of noise than the standard imaging. As we show in Figure  \ref{s9} , the relative decrease of $\zeta$ due to the background noise is the highest, when only mean signal is used in the estimation. This result was to be expected, as the authors of SOFI method claim, that the background noise reduction is its important feature \cite{Dertinger2009}. Nevertheless, the background noise affects the performance of each of the analyzed methods. 

\subsection{Non-equal states probabilities}
\label{secasym}
\begin{figure}[ht]

\begin{subfigure}{.45\linewidth}
  \centering
  \includegraphics[width=\linewidth]{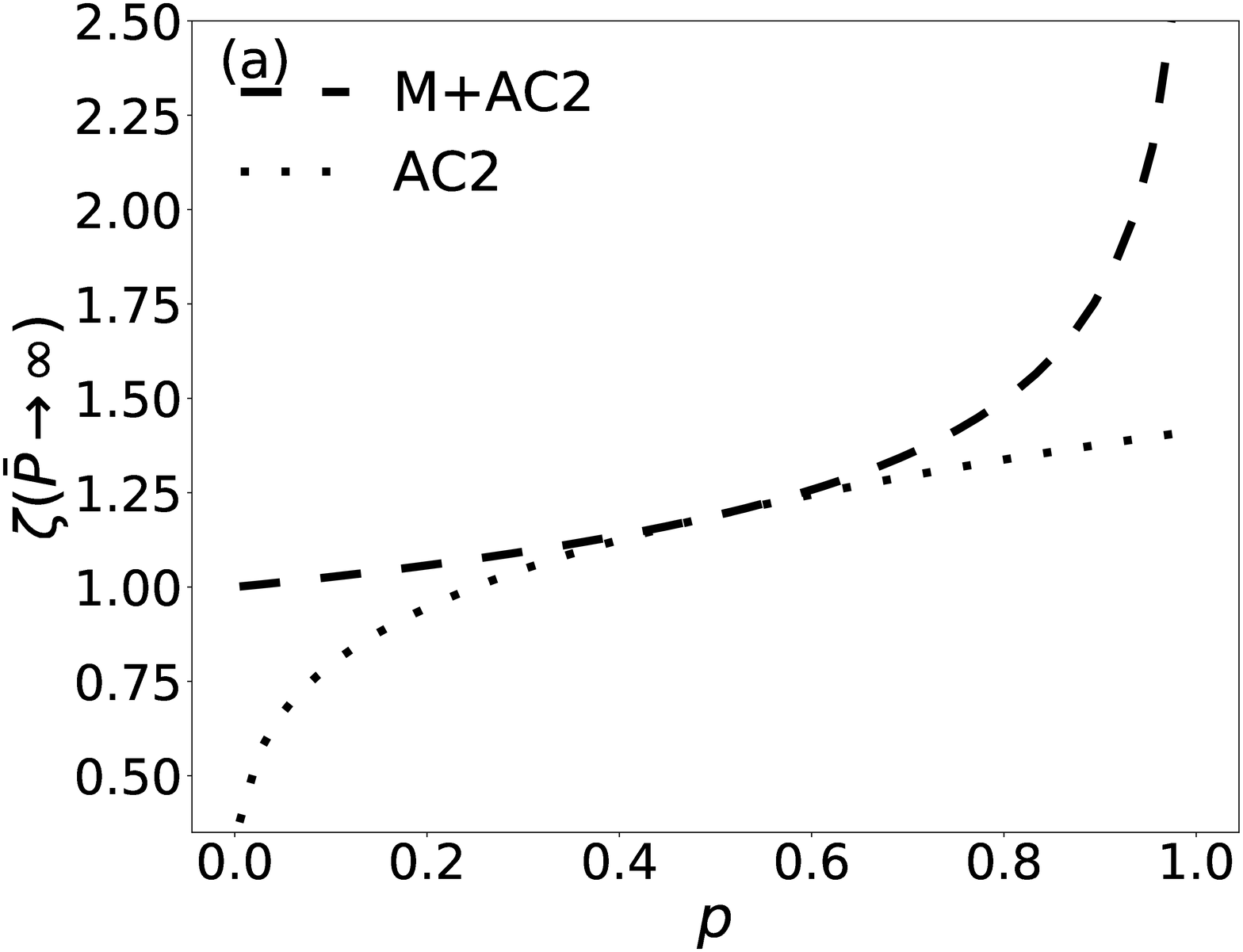}
\end{subfigure}%
\begin{subfigure}{.45\linewidth}
  \centering
  \includegraphics[width=\linewidth]{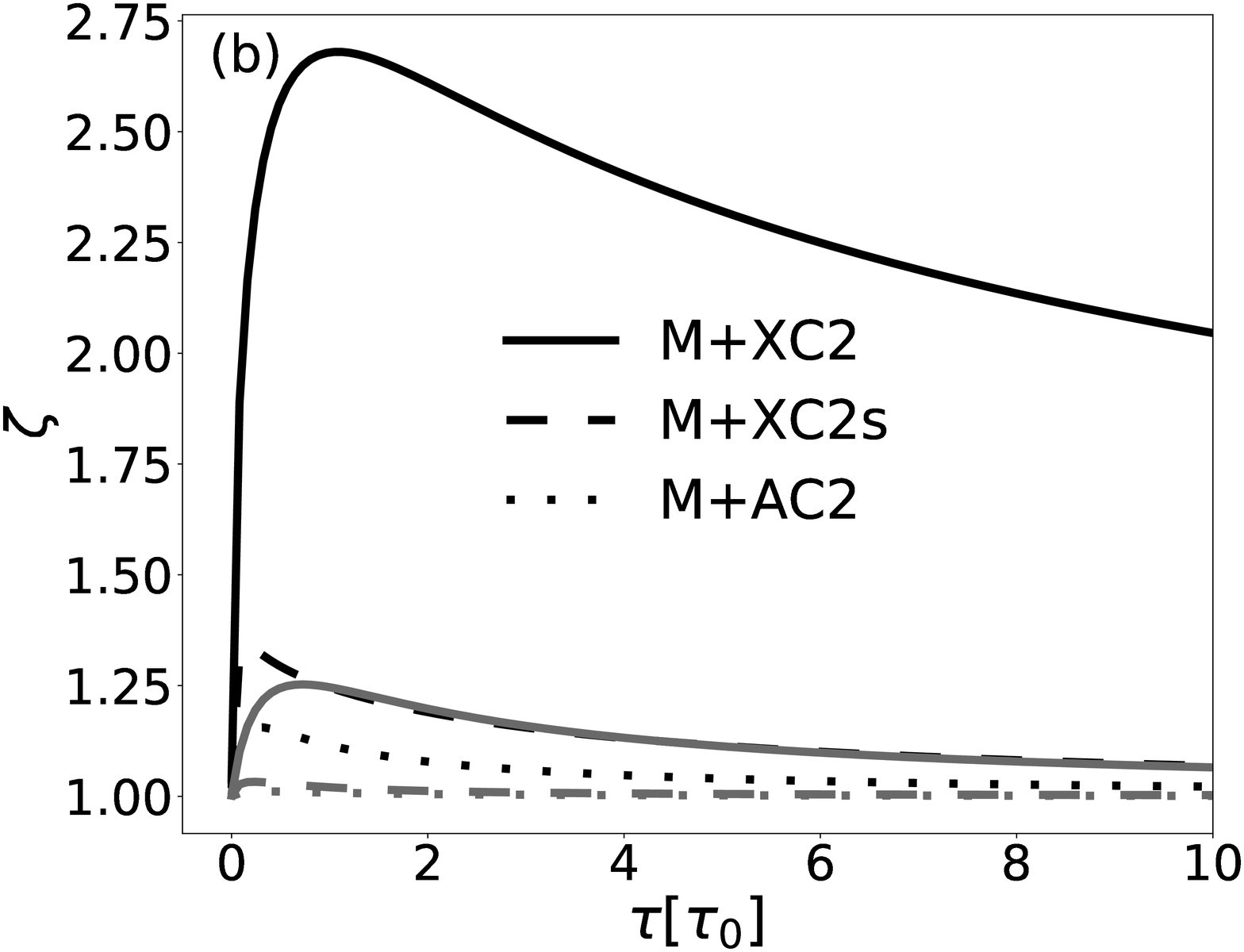}
\end{subfigure}

\caption{ The RGL in the limit of infinitely bright sources and strong fluctuations ($\alpha=1$) as a function of $p= P(q_\textrm{off})$ is sketched in (a) (the simplified blinking model is used). One can observe, that estimation schemes based on (AC2) and (M+AC2) are only equivalent for $p=0.5$. For $p>0.5$ the RGL can be larger than $\sqrt[4]{2}$, but the (AC2) scheme never allows to beat $\zeta<\sqrt{2}$ limit. The RGLs computed for the realistic model are sketched in (b). Gray lines denote $\tau_\textrm{off}=0.4 \tau_0$, $\tau_\textrm{on}=1.6 \tau_0$ case, whereas  $\tau_\textrm{off}=1.6 \tau_0$, $\tau_\textrm{on}=0.4 \tau_0$ for black lines, and $\alpha=1$ for both cases. A significant increase of $\zeta$ is possible, when off- state is favored. }
\label{f56}

\end{figure}

So far, we have presented results for $\tau_\textrm{on} = \tau_\textrm{off}$ (realistic model) or $p=1/2$ (simplified model). However, it's known that real emitters sometimes break this assumption, and favor one of the states.  One can  observe, that if off- state is more probable, the RGL becomes higher because more photons are emitted within frames in which two sources have different brightness. One should also take this asymmetry into account while dealing with optimizing the time frame for different estimation schemes---see Figure ~\ref{f56}.
\subsection{Weighted summation of cross-cumulants}
\label{sec3}
Let us examine the performance of an improved version of the image reconstruction scheme based on summing of the covariances of pixel pairs with the same centroid (M+XC2s).
The authors of \cite{vandenberg2016model} propose to add weights to the summation procedure in order to maximize the SNR of the reconstructed image. Let $\kappa_1, \kappa_2,...\kappa_k$ be the covariances contributing to the same centroid. The signal located in this centroid is equal to $S= \kappa_1 + \kappa_2 + ... +\kappa_k$, when the basic scheme is considered. More generally, one can use the formula $S=w_1 \kappa_1 + w_2 \kappa_2 +...+w_k \kappa_k$, where $w_i$ are some arbitrary weights. As shown in \cite{vandenberg2016model}, in order to maximize the $\textrm{SNR} = \frac{\left< S \right>}{\sqrt{\textrm{Var}(S)}}$, one should choose $w_i$ satisfying the following conditions:
\begin{equation}
\label{weights}
    \sum_{i=1}^k w_i \left( \frac{A_{mi}}{\kappa_m} - \frac{A_{1i}}{\kappa_1}\right) = 0,
\end{equation}
where $m \in \{ 2,3,...,k \}$, and $A_{ij} = \textrm{cov}(\kappa_i, \kappa_j)$. In reality, one obviously has no access to the exact values of $\kappa_i$ and $A_{ij}$, so their estimators must be inserted into \eqref{weights} in order to obtain the optimal weights. However, in the limit of large number of collected frames,  the difference between the actual values and the estimators doesn't affect the weights $w_i$ significantly. We will therefore assume, that the exact values of $\kappa_i$ and $A_{ij}$ are used in the weights computing procedure.

We can assume that $w_1=1$ without the loss of generality, as multiplying each weight by the same factor doesn't change the information content of the computed linear combination (in particular, the SNR remains the same). It's then straightforward to compute the rest of weights, by solving the set of equations \eqref{weights}. Then, the procedure for calculating $\bm \mu$, $\bm \Sigma$ , and afterwards $\mathcal F$ and $\zeta$, is the same as the one described in \ref{secC}---one only needs to replace the sum of the covariances with the proper linear combination. 

\begin{figure}[ht]

\begin{subfigure}{.45\linewidth}
  \centering
  \includegraphics[width=\linewidth]{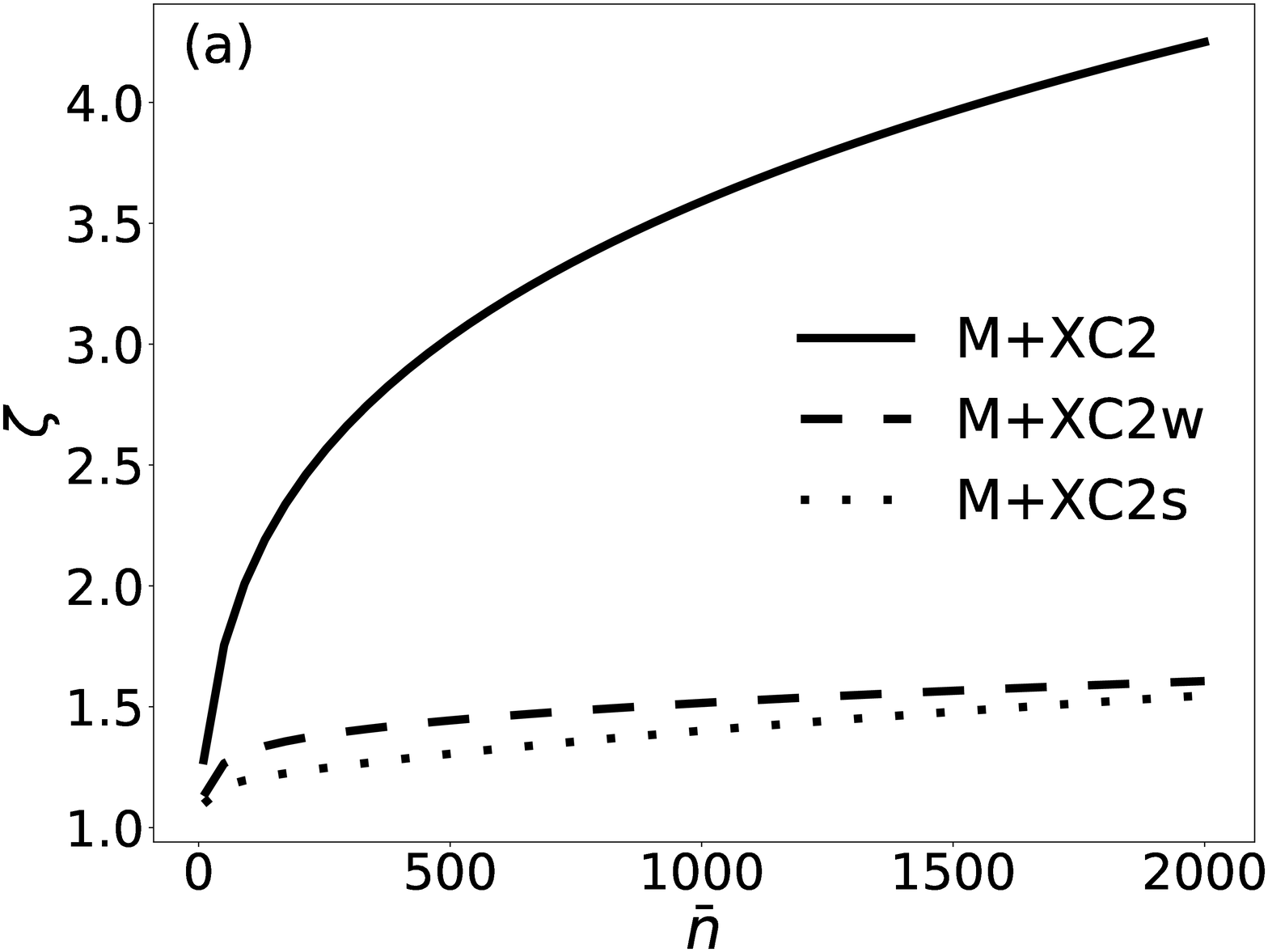}
\end{subfigure}%
\begin{subfigure}{.45\linewidth}
  \centering
  \includegraphics[width=\linewidth]{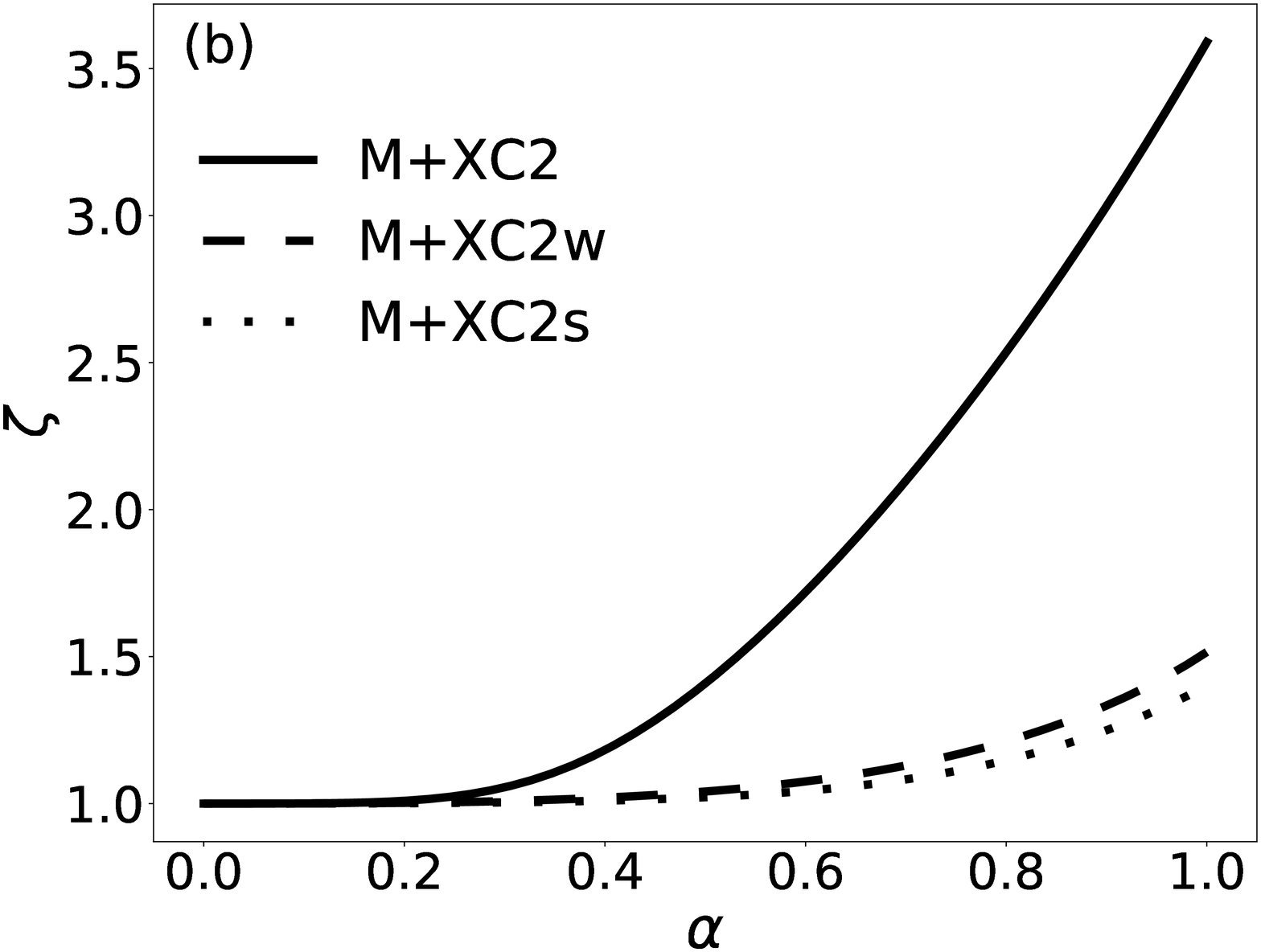}
\end{subfigure}

\caption{ $\zeta$ as a function of $\bar{n} = \bar P \tau$ (for $\alpha=1$) (a), and as a function of $\alpha$ (for $\bar{n} = 1000$)  (b), for different estimation schemes. The improved scheme, based on weighted sums of covariances (M+XC2w), outperforms the basic scheme (M+XC2s) only slightly. The simplified model (independent frames) was used, with parameters $p=0.5$, $\Delta x = 0.5 \sigma$.}
\label{f78}

\end{figure}

As we show in Figure ~\ref{f78}, the improved procedure (M+XC2w) allows only a slight increase of $\zeta$ compared to the previously examined summation scheme (M+XC2s). The limit dictated by the approach in which covariances are treated as independent observables (M+XC2) is still far from being achieved. We therefore claim, that the problem of finding the optimal way to reconstruct the image using covariances, still remains open. It's partially due to the fact, that the SNR itself doesn't provide a full description of noise, when noise is correlated, as in the analyzed schemes.

\subsection{Non-classical sources and 3D imaging}
\label{secantb}
The photon number fluctuations utilized in SOFI are described by super-Poissonian distribution, and therefore can be explained with the help of semi-classical models. However, sub-Poissonian photon number distributions can be used to obtain super-resolution as well.  As it's believed (and justified by the effective PSF analysis) \cite{schwartz2013superresolution}, the resolution can be increased by a factor $\sqrt 2$ if two emitters are imaged, two-photon frames are observed, and due to anti-bunching phenomenon one can be sure, that at most one photon can be emitted from a single source within a single frame. Let's challenge this statement, and compute $\zeta$ for the described case and check how the fundamental shot noise affects the possibility of obtaining super-resolution. The PDF of measuring a frame consisting of photons at position $x_1, x_2$ is
\begin{equation}
    p_\theta(x_1, x_2) = \frac{  U (x_1+\theta/2) U(x_2 - \theta/2) + U(x_2+\theta/2) U(x_1 - \theta/2) }{2}.
\end{equation}
After substituting the Gaussian form of $U$, expanding $p_\theta(x_1,x_2)$ into series around $\theta=0$, and using \eqref{Fisher}, we obtain the formula for the two-photon frame FI:
\begin{equation}
    \mathcal F_{(2)} (\theta) = \frac{\theta ^2}{2 \sigma^4} + \mathcal O(\theta^4/\sigma^6)
\end{equation}
The FI per one photon in this case is equal to $\mathcal F = \frac{1}{2} \mathcal F_{(2)}$, which after using \eqref{MRG} leads to $\zeta = \sqrt[4]{2}$. Despite optimistic assumptions (no noise apart from shot noise, only two-photon frames), our RGL again turns out to be lower than the resolution gain predicted by the PSF analysis.

The RGL introduced in this paper only applies to the transverse resolution. It is however known, that SOFI is a 3D method, and as such, allows to improve axial resolution as well. Nevertheless, the most important aspects of  SOFI, which we want to study (i. e. how noise affects obtainable resolution) are all well illustrated with our model, in which both sources lie precisely in the image plane. To extend the whole reasoning to the 3D case, one should proceed analogously as in \cite{zhou2019quantum3d}. The information about the axial separation between the sources is encoded in the size of resulting PSFs, which are affected by the deviation from the image plane. When the deviation is very small, the FI associated with its estimation vanishes because the size of the PSF changes slowly with the axial displacement in this region. As with transverse resolution, the fluctuations will not result in a non-vanishing FI, but will rather result in a larger values of FI for small, yet non-zero deviations. 
\section{Summary}
To summarize, we have provided a quantitative approach based on estimation theory, to compute performance limits on super-resolution imaging methods that utilize sources brightness fluctuations. By focusing on the rudimentary problem of resolving two point sources, we were able to provide a single meaningful quantity that allows to compare resolution gain of different methods and identify the optimal detection frame time. The study, has on one hand identified new fundamental limitations of some of the methods (e.g. 2nd auto-cumulant method), as well as indicated space for improvement of other methods (e.g. cross-cumulant based methods). Even though the study was based on a two point sources imaging problem, the obtained resolution limits can be regarded as 
valid, but possibly not tight, limits also in more complex multiple sources imaging case. 
In order to obtain tighter bounds, a more detailed quantitative study would be required, invoking the concepts of multi-parameter estimation theory. There are a number of ways 
how to phrase a complex imaging model as a multi-parameter estimation problem, and  apart from the most obvious pixel by pixel image parametrization, may include focusing on moments of spatial intensity distribution \cite{Zhou2019, Chrostowski2017} or
its Fourier components \cite{koppell2020}.  
 Generalization of such studies to the case of sources with fluctuating brightness, while non-trivial, seems possible and may provide a further insight into the potential as well as the limitations of SOFI and related methods.

\section*{Acknowledgments}
We thank Konrad Banaszek for fruitful discussions. We acknowledge support from the National Science Center (Poland) grant No. 2016/22/E/ST2/00559.

\appendix

\section{Details of Fisher Information computations}
\label{sec1}
A detailed derivation of the FI associated with the estimation of the distance between two point sources in different scenarios will be provided in this section. We will start with the simplest, well known case of non-fluctuating Poissonian sources to justify \eqref{FSI} from the main text. Afterwards, intensity fluctuations will be added to our scheme. Simplified fluctuations model, in which subsequent frames are independent will be examined. We are going to derive a very general formula for FI, which is suitable for different types of intensity fluctuations, not only for two-level emitters presented in the main text. Then, adequate simplifications will be made to obtain the formula for $\zeta_\textrm{max}$ (\eqref{zetamax1}, \eqref{zetamax2}, \eqref{zetamax3}). In the last part of this section, calculations of the FI associated with different cumulant based algorithms, for both simplified, and realistic Markov-process based model, will be described.
\newline \newline
Consider two point emitters placed at $-\theta/2$ and $\theta/2$. The PSF of the imaging system is assumed to be Gaussian with a standard deviation $\sigma$: 
\begin{equation}
\label{PSF}
U(x) = \left(2 \pi \sigma^2 \right)^{-1/2} \exp \left( -\frac{x^2}{2 \sigma^2} \right).
\end{equation}
Our goal is to compute the FI per one photon $\mathcal F (\theta)$. To do so, one needs to compute the FI for the whole measurement $\mathcal F_\textrm{meas}(\theta)$ and then divide it by the average total number of photons. Our task becomes slightly easier if the whole measurement output can be divided into independent, identically distributed parts (e.g. intensities measured in different, independent frames). It's then enough to compute the FI associated with only one of such independent parts because FI is additive for independent random variables.
\subsection{Non-fluctuating emitters}
\label{secA}
This case is particularly easy because subsequent photons are not correlated, and the FI per one photon can be calculated directly. Sources are equally bright, and the spatial resolution of the detector is infinite. Each photon position $x$ is independently drawn from the probability density function (PDF):
\begin{equation}
    p_\theta (x) = \frac{U \left(x+\theta/2 \right) + U \left( x-\theta/2 \right)}{2}.
\end{equation}
Now the FI can be computed with the help of \eqref{Fisher} in which vector $\bm N$ consists of just one element---a detected photon position $x$. The $\mathcal F$ can be therefore expressed as an integral
\begin{equation}
    \mathcal F (\theta) = \int_{- \infty}^{\infty} \frac{1}{p_\theta(x)}  \left( \frac{ \partial p_\theta(x)}{ \partial \theta} \right)^2 \, \textrm{d}  x,
\end{equation}
which after substituting the Gaussian form of $U(x)$ simplifies to 
\begin{equation}
    \sigma^2 \mathcal F (\theta) = \frac{1}{4} - \int^{\infty}_{-\infty} \frac{x^2 \exp\left(-\frac{1}{8 \sigma^2} (\theta -2 x)^2\right)}{2 \sigma^3 \sqrt{2 \pi } \left( \exp \left(\theta x/\sigma^2 \right)+1\right)}\, \textrm{d} x.
\end{equation}
To obtain the analytical form of the above integral for $\theta \ll \sigma$, one can expand the integrated function in the series around $\theta=0$, and perform the integration term by term to conclude that
\begin{equation}
\label{Fclass}
    \mathcal F (\theta) = \frac{1}{\sigma^2} \left( \frac{\theta^2}{8 \sigma^2}-\frac{\theta^4}{16 \sigma^4}+\frac{\theta^6}{24 \sigma^6}+... \right),
\end{equation}
which is consistent with \eqref{FSI}. In order to obtain the values of $\mathcal F(\theta)$ for larger $\theta$, the introduced integral must be calculated numerically. To compute $\mathcal F$ in case of non-zero pixel size $\Delta x$, one needs to construct vector $\bm N$ which consists of mean values of the signal in different pixels only, and then proceed as in \ref{secC}.
\subsection{Fluctuating emitters, independent frames}
\label{secB}
For the rest of this section, the assumption $\sigma = 1$ will be made.
Let's consider the simplified model of fluctuations which is slightly more general than the one described in the main text. In each independent frame relative brightness of emitters placed at $-\theta/2$ and $\theta/2$, denoted by $q_1$ and $q_2$ respectively, is independently drawn from the same arbitrary probability distribution $P(q_i)$. The frame time and the mean emitters power are denoted by $\tau$ and $\bar P$ respectively---for the sake of simplicity we are going to use the quantity $\bar{n} = \bar P \tau$, which is the only relevant quantity as long as frames are independent, and is proportional to the mean number of photons detected per frame.

\begin{figure}[ht]

\begin{subfigure}{.5\linewidth}
  \centering
  \includegraphics[width=\linewidth]{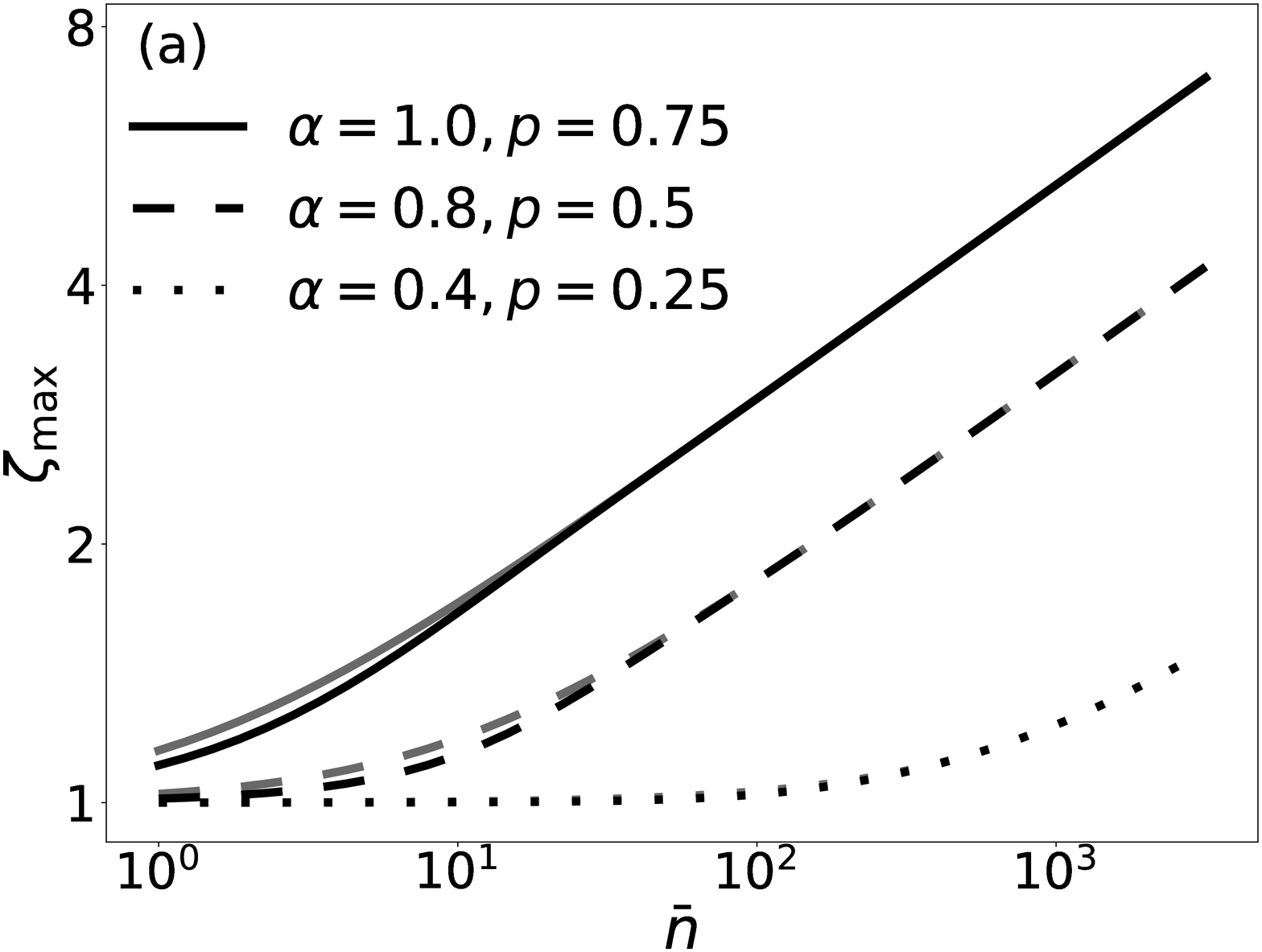}
  \label{f41}
\end{subfigure}%
\begin{subfigure}{.5\linewidth}
  \centering
  \includegraphics[width=\linewidth]{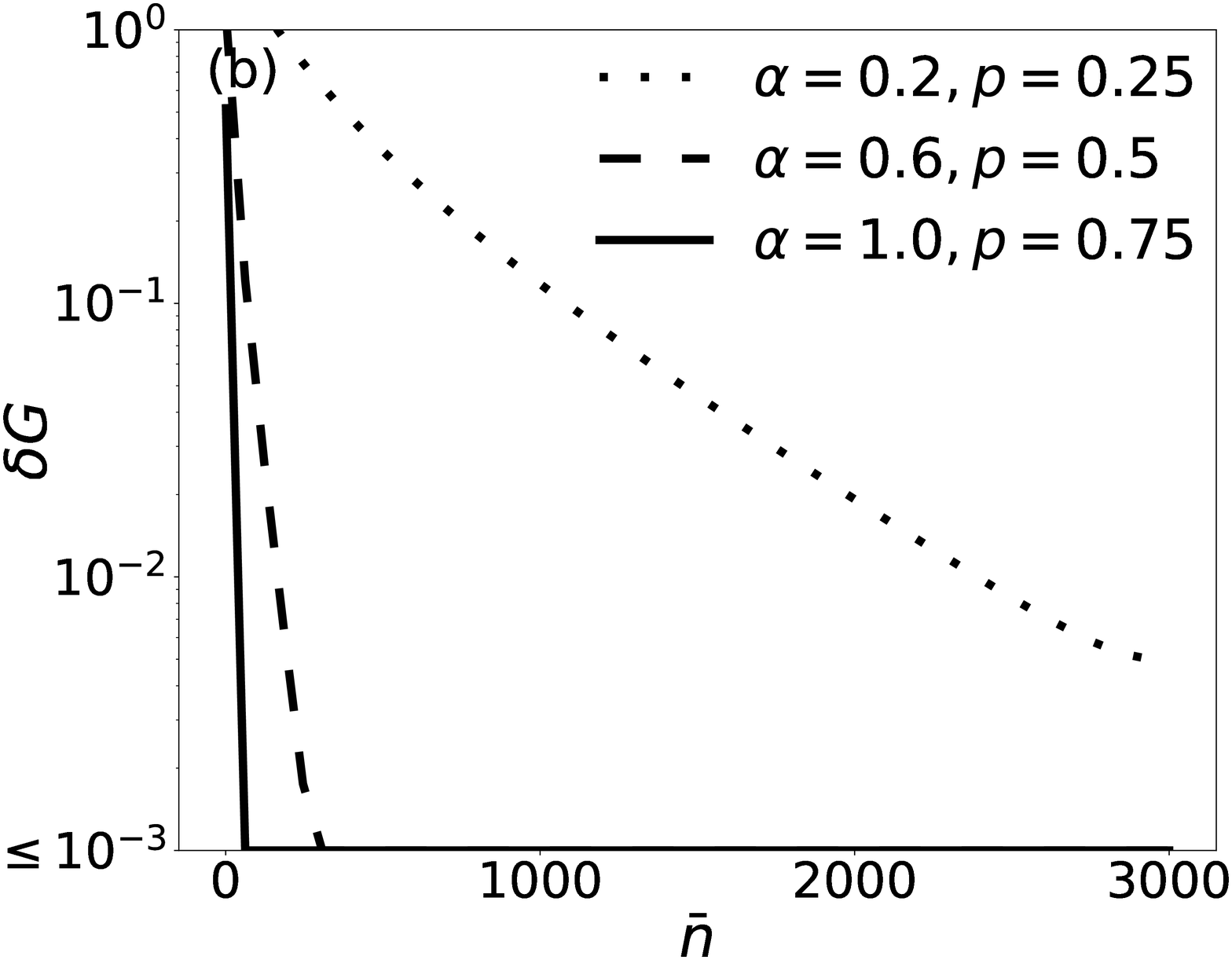}
  \label{f42}
\end{subfigure}
\caption{The comparison between $\zeta (\bar{n})$ computed numerically (black lines), and its analytical approximation (gray lines) for different $p$ and $\alpha$ values is shown in (a). The relative error  $\delta G = \frac{G(p,\alpha, \infty) - G(p, \alpha, \bar{n})}{G(p,\alpha, \bar{n})}$ as a function of $\bar{n}$ is sketched in (b).}
\label{f4}
\end{figure}

As mentioned in the main text, in order to compute the FI per one photon $\mathcal F$, we compute the FI for each fixed number of photons in a frame separately  ($\mathcal F_{(n)}$), and then use the formula
\begin{equation}
\label{F1}
 \mathcal F = \frac{\left< \mathcal F_{(n)} \right>_n}{\left< n\right>_n}.   
\end{equation}
 $\mathcal F_{(n)}$ is the FI associated with the conditional PDF 
\begin{equation}
\label{S8}
    p_\theta(x_1,...,x_n|n) = \int \textrm{d} q_1 \int \textrm{d} q_2\, p_\theta(x_1,...x_n|n,q_1,q_2) P(q_1,q_2|n).
\end{equation}
Notice, that the above equation is a generalised form of \eqref{12}. Equations \eqref{10}, \eqref{11}, \eqref{13} are still valid in the general case, and can be used to compute conditional probabilities $p_\theta(x_1,...,x_n|q_1,q_2,n)$ and $p_\theta(x_1,...,x_n|n)$.
 We are now going to find an explicit expression for $p_\theta(x_1,...,x_n|n)$ to compute $\mathcal F_{(n)}$ directly from the definition of FI. To do so, let's begin with inserting \eqref{11} and \eqref{PSF} into \eqref{10}. Before performing the product in \eqref{10}, we expand each factor into series around $\theta=0$. After keeping only the leading terms, we obtain
\begin{equation}
\label{pxkq}
    p_{\theta}(x_1, ..., x_n | n, q_1, q_2) = (2 \pi)^{-n/2} \left( \prod_{i=1}^n  e^{-\frac{1}{8} x_i^2} \right) \left( 1+ A_1 \theta+ A_2 \theta^2 + A_3 \theta^3+ A_4 \theta^4+... \right),
\end{equation}
where

\begin{equation} \label{A2}
A_2= \frac{1}{8} \sum_{i=1}^n x_i^2 + \frac{Q_2}{4} \sum_{i<j} x_i x_j - \frac{1}{8}n ,
\end{equation}

\begin{multline} \label{A4}
A_4=\frac{1}{384} \sum_{i=1}^n x_i^4 - \frac{n}{64} \sum_{i=1}^n x_i^2 + \frac{1}{128}n^2 + \frac{Q_2}{96} \sum_{ i \neq j} x_i(x_j^3 - 3x_j) + \frac{1}{64} \sum_{i<j} x_i^2 x_j^2 +\\+ \frac{Q_2}{32} \sum_{i<j,k \neq i, k \neq j} x_i x_j (x_k^2 - 1) + \frac{Q_4}{16} \sum_{i<j<k<m} x_i x_j x_k x_m ,
\end{multline}
and the quantity $Q_k$ is defined as
\begin{equation}
Q_k \equiv \left( \frac{q_1-q_2}{q_1+q_2} \right)^k.
\end{equation}
We don't specify the form of $A_1$ and $A_3$, which are not relevant as will be argued below. Now we can use \eqref{S8} and \eqref{pxkq} to obtain $ p_\theta(x_1,...,x_n|n)$. Let us denote the expected value of a function $X(q_1, q_2)$ with respect to $P(q_1, q_2 |n)$ by $\left< X \right>_{q|n}$:
\begin{equation}
\left< X \right>_{q|n} \equiv \int X(q_1, q_2) P(q_1, q_2 |n) dq_1 dq_2.
\end{equation}
 Notice, that if we replace all $Q_k$ terms in \eqref{A2} and  \eqref{A4} by their mean values $\left< Q_k \right>_{q|n}$, we obtain the mean values of coefficients---$\left< A_2 \right>_{q|n}$ and $\left< A_4 \right>_{q|n}$. Furthermore:
 \begin{equation}
 \label{pxk}
 p_\theta(x_1,...,x_n|n) = (2 \pi)^{-n/2} \left( \prod_{i=1}^n  e^{-\frac{1}{8} x_i^2} \right) \left( 1+ \left< A_2 \right>_{q|n} \theta^2 + \left< A_4 \right>_{q|n} \theta^4 + \mathcal O (\theta^6) \right).
 \end{equation}
 We have just used the fact that odd coefficients $A_1,A_3,A_5,...$ contain only terms proportional to $Q_l$, where $l$ is odd. Moreover, from statistical identity of both sources it follows that $\left< Q_l \right>_{q|n}=0$ for odd $l$---that's the reason why odd coefficient could have been neglected from the beginning, and only even $\theta$ powers are present in \eqref{pxk}. We are now going to calculate $\mathcal F_{(n)}$ using \eqref{Fisher} which takes the form
 \begin{equation}
    \mathcal F_{(n)}  = \int \frac{1}{p_\theta(x_1,...,x_n|n)}  \left( \frac{ \partial p_\theta(x_1,...x_n|n)}{ \partial \theta} \right)^2  \textrm{d}  x_1 ... \textrm{d} x_n,
\end{equation}
which simplifies to
\begin{equation}
    \mathcal F_{(n)} = (2 \pi)^{-n/2} \int  \prod_{i=1}^n  e^{-\frac{1}{8} x_i^2}  \left(  4 \left< A_2 \right>_{q|n}^2 \theta^2 + \left(16 \left< A_2 \right>_{q|n} \left< A_4 \right>_{q|n} - 4 \left< A_2 \right>_{q|n}^3 \right) \theta^4 + \mathcal O (\theta^6) \right)  \textrm{d}x_1... \textrm{d}x_n .
\end{equation}
After inserting the formulas for $\left< A_2 \right>_{q|n}$ and $\left< A_4 \right>_{q|n}$, and performing the integration, we obtain
\begin{equation}
\label{Fkfull}
    \mathcal F_{(n)} = \theta^2 \left[ \frac{n}{8} + \frac{\left< Q_2 \right>_{q|n}^2}{8}  n(n-1)   \right] + \theta^4 \left[ -\frac{n}{16} - \frac{\left< Q_2 \right>_{q|n}^2}{16} n (n-1) -\frac{\left< Q_2 \right>_{q|n}^3}{16} n(n-1)(n-2) \right] + \mathcal O (\theta^6).
\end{equation}
Notice, that now $\theta \ll 1$ condition is not sufficient to ensure that the term with $\theta^4$  is negligible compared to the $\theta^2$ term, because the powers of $n$ are different in both terms. The approximation
\begin{equation}
    \mathcal F_{(n)} \simeq  \theta^2 \left[ \frac{n}{8} + \frac{\left< Q_2 \right>_{q|n}^2}{8}  n(n-1)   \right]
\end{equation} 
 is nevertheless justified provided $\theta^4 n^3 \ll \theta^2 n^2$, which is equivalent to the condition
 $\theta \ll n^{-1/2}$. The above inequality holds for all $n$ that give relevant contribution to the final result if $\theta \ll \bar{n}^{-1/2}$.  Equation \eqref{F1} allows us to compute the one-photon FI:
 \begin{equation}
     \mathcal F = \frac{\theta^2}{8} \left[ 1 + \frac{\left< \left< Q_2 \right>_{q|n}^2  n(n-1) \right>_n}{  \left< n \right>_n} \right] + \mathcal O (\theta ^4 \bar{n} ^2).
 \end{equation}
 Note that if emitters don't fluctuate, $q_1$ is always equal to $q_2$, so $\left< Q_2 \right>_{q|n} =0$, and we recover \eqref{Fclass} up to the 2nd order of $\theta$---using \eqref{Fkfull} one can  additionally check that the coefficient at $\theta^4$ for non-fluctuating case is also correctly retrieved. The quantity $\left< Q_2 \right>_{q|n}$ can be regarded as a measure of fluctuations intensity---it becomes larger, if the normalized difference between $q_1$ and $q_2$ takes large values with high probability. As intuitively expected, the FI per one photon increases with $\left< Q_2 \right>_{q|n}$, as well as RGL defined in \eqref{MRG}, which in our case has a form
 \begin{equation} \label{zetamax}
     \zeta_\textrm{max}=\left(1 + \frac{\left< \left< Q_2 \right>_{q|n}^2  n(n-1) \right>_n}{  \left< n \right>_n}\right)^{1/4}.
 \end{equation}
 In order to obtain a more specific expression for $\zeta_\textrm{max}$, let's consider the two-level model of emitters mentioned in the main text, i.e.
   \begin{align}
   &P \left(q_i = q_\textrm{off} \right) = p \\
   &P\left(q_i = q_\textrm{on} \right) = 1-p
   \end{align}
for $i \in \{1,2\}$. Recall that $q_\textrm{off}+q_\textrm{on}=1$, and the fluctuation strength is defined as $\alpha=1-q_\textrm{off}/q_\textrm{on}$. It's easy to show that the mean number of photons detected per frame is 
\begin{equation}
\label{kmean}
    \left< n \right>_n = 2 \bar{n} \left( p^2 q_\textrm{off} + p(1-p) + (1-p)^2 q_\textrm{on} \right).
\end{equation}
 The variable $Q_2$ takes a non-zero value only in two equally probable cases when $q_1 \ne q_2$, so
 \begin{equation}
 \label{S26}
      \left< Q_2 \right>_{q|n} = 2 \left( q_\textrm{on} - q_\textrm{off} \right)^2 P(q_1=q_\textrm{on}, q_2=q_\textrm{off}|n).  
 \end{equation}
 The probability $P(q_1=q_\textrm{on},q_2=q_\textrm{off}|n)$ is calculated using \eqref{13}, and the fact that 
 \begin{equation}
 P(n|q_1, q_2) = \frac{(\bar{n} (q_1+q_2))^n \exp(-\bar{n}(q_1+q_2))}{n!}. 
 \end{equation}
The expression present in \eqref{zetamax} can be written as an infinite sum
\begin{equation}
\label{q2qk}
    \left< \left< Q_2 \right>_{q|n}^2  n(n-1) \right>_n=\sum_{n=0}^{\infty} \left< Q_2 \right>_{q|n}^2  n(n-1) P(n),
\end{equation}
 which after inserting \eqref{S26} and making some simplifications (e.g.  changing variables $q_\textrm{on}, q_\textrm{off}$ to $\alpha$) takes the form
\begin{equation}
\label{S29}
  \left< \left< Q_2 \right>_{q|n}^2  n(n-1) \right>_n = 4 p^2 (1-p)^2 \left(\frac{\alpha}{2-\alpha} \right)^4 \bar{n}^2  S,
\end{equation}
 where
 \begin{equation}
S = \sum_{n=0}^{\infty} \frac{e^{-\bar{n}} \bar{n} ^n}{n!} \left( B e^{A \bar{n}}(1-A)^n+C+D e^{-A \bar{n}} (1+A)^n\right)^{-1},
 \end{equation}
and the following definitions are used: $A = \frac{\alpha}{2-\alpha}$, $B=p^2 (1-A)^2$, $C=2p(1-p)$,  $D=(1-p)^2(1+A)^2$. 
 Joining together \eqref{S29},~(\ref{kmean}), and (\ref{zetamax}), we obtain the following expression:
\begin{equation}
\label{S31}
    \zeta_\textrm{max}=\left(1 + G(p,\alpha, \bar{n}) \bar{n} \right)^{1/4},
\end{equation}
where 
\begin{equation}
\label{gdef}
    G(p,\alpha, \bar{n})= \frac{2 p^2 (1-p)^2 \alpha^4}{ (2-\alpha)^3 (1-p \alpha)} S. 
\end{equation}
To obtain the value of $\zeta_\textrm{max}$ for arbitrary parameters, one needs to approximate the infinite sum $S$ numerically. However, it's possible to prove that (see \ref{secD})
\begin{equation}
\label{S32}
    \lim_{\bar{n} \rightarrow \infty} S = \left\{ \begin{array}{ll}
C^{-1}& 0<\alpha \le 1 \\
(B+C+D)^{-1} & \alpha=0
 \end{array} \right. ~ ,
\end{equation}
which allows us to provide an analytical expression for $\zeta_\textrm{max}$ scaling in the infinitely bright sources regime:
\begin{equation}
    \lim_{\bar{n} \rightarrow \infty} \zeta_\textrm{max} \bar{n} ^{-1/4} = \left(\frac{p(1-p) \alpha^4}{(2-\alpha)^3(1-p \alpha)} \right)^{1/4}.
\end{equation}
Numerical analysis show, that the replacement of $S$ by $C^{-1}$ in \eqref{S31}, which leads to equation
\begin{equation}
\label{S34}
\zeta_\textrm{max} \simeq \left( 1 + \frac{p(1-p) \alpha^4}{(2-\alpha)^3(1-p \alpha)} \bar{n} \right)^{1/4}  ,
\end{equation}
becomes a good approximation for large enough $\bar{n}$. The comparison between $\zeta_\textrm{max}$ computed numerically for finite $\bar{n}$ using \eqref{S31} and its analytical approximation valid for large $\bar{n}$ \eqref{S34} is shown in Figure  \ref{f4}.

\subsection{Cumulant based algorithms}

\label{secC}
From now on the spatial resolution of the camera is not assumed to be infinite, and the whole detection area is divided into $M_\textrm{pix}$ pixels of size $\Delta x$. The positions of the centroids of subsequent pixels are denoted by $x_1,...,x_{M_\textrm{pix}}$. The detection time is divided into $M_\textrm{fr}$ frames, $m$-th  frame covers the time interval $\left[ (m-1) \tau, m \tau  \right]$, where $m \in \{ 1,2,...,M_\textrm{fr} \}$. We are going to consider only two-level blinking model, and stay with its simplified version with independent frames for a while. Our goal is to compute the FI associated with different choices of vector $\bm N$. Let's consider the case studied in the main text, in which $\bm N$ is described by \eqref{6}. Subsequent frames are independent, so the central limit theorem can be directly used to prove that $\bm N$ is normally distributed. 

One needs to use slightly more subtle arguments to extend the above reasoning to the estimation based on 2nd auto-cumulant only (AC2). $\bm N$ consists of 2nd auto-cumulant (variance) estimators for each pixel 
\begin{equation}
\bm v_m = \left[n_{1,m}^2 - \left<  n_1 \right>^2, \cdots, n_{M_\textrm{pix},m}^2 - \left<  n_{M_\textrm{pix}} \right>^2  \right]^T,
\end{equation}
where $ \left< n_i \right> = \frac{1}{M_\textrm{fr}}\sum_{m=1}^{M_\textrm{fr}} n_{i,m}$ denotes the mean value estimator. 
Unfortunately, this estimator depends on detected photon numbers from different frames, so vectors $\bm v_m$ are not mutually independent anymore. However, in the limit $M_\textrm{fr} \rightarrow \infty$ the mean value estimator becomes very accurate compared to the variability of the number of photons in a given pixel in a single frame because the variance of $n_{i,m}$ doesn't depend on $M_\textrm{fr}$, and the variance of $\left< n_i \right>$ scales as $1/M_\textrm{fr}$. That means, that the replacement of the mean value estimator with its exact value in  $\bm v_m$ doesn't affect the distribution of $\bm N$ in the limit $M_\textrm{fr} \rightarrow \infty$.  Vectors $\bm v_m$ become independent after making the described replacement, which allows us to conclude, that $\bm N $ is normally distributed.

In order to compute the FI associated with the normally distributed vector $\bm N $ in the most general case in which both the mean vector $ \bm\mu$, and the covariance matrix $ \bm\Sigma$ depend on the estimated parameter $\theta$, one can use the formula \cite{Kay1993}
\begin{equation}
\label{S36}
\mathcal F_\textrm{(meas)} = \frac{\partial \bm \mu^\top}{\partial \theta} \bm \Sigma^{-1} \frac{\partial \bm \mu}{\partial \theta} + \frac{1}{2} \textrm{tr} \left( \bm \Sigma^{-1} \frac{\partial \bm \Sigma}{\partial \theta} \bm \Sigma^{-1} \frac{\partial \bm \Sigma}{\partial \theta} \right).
\end{equation}
If vectors $\bm v_m$ are independent and identically distributed (i.i.d.), the mean vector and the covariance matrix for each $\bm v_m$ are denoted by $\bm \mu_{(1)}$ and $\bm \Sigma_{(1)}$, then $\bm \mu = \bm \mu_{(1)}$, and $ \bm \Sigma = \frac{1}{M_\textrm{fr}} \bm \Sigma_{(1)}$ .  We therefore see, that the 2nd term in \eqref{S36}  is neglibible compared to the 1st term in the limit $M_\textrm{fr} \rightarrow \infty$ (provided the first term is non-zero), and hence in this limit we may write
\begin{equation}
\label{S37}
\mathcal F_\textrm{(meas)} = \frac{\partial \bm \mu^\top}{\partial \theta} \bm \Sigma^{-1} \frac{\partial \bm \mu}{\partial \theta} =M_\textrm{fr} \frac{\partial \bm \mu_{(1)}^\top}{\partial \theta} \bm \Sigma_{(1)}^{-1} \frac{\partial \bm \mu_{(1)}}{\partial \theta}.
\end{equation}In order to compute the FI per one photon $\mathcal F$ one needs to compute the elements of $\bm \mu $ and $\bm \Sigma$, use \eqref{S37}, and then divide $\mathcal F _\textrm{(meas)}$ by the average total photon number. Let $v_{1,m}, v_{2,m},...,v_{n,m}$ be the elements of the vector $\bm v_m$. In some cases we are going to use a short-hand notation $v_{i,1} \equiv v_i, n_{i,1} \equiv n_i$, because the 2nd index can be omitted in many situations when single frame statistics are considered. Then:
\begin{equation}
\label{mu}
\bm \mu = \left[  \left<v_1 \right>, \left<v_2 \right>, ...,\left< v_n \right>\right]^T,
\end{equation}
\begin{equation}
\bm \Sigma = \frac{1}{M_\textrm{fr}} \left( \left[ \begin{array}{cccc} \left< v_1 v_1 \right> & \left<v_1 v_2 \right> & \ldots & \left< v_1 v_n \right>  \\ \left< v_2 v_1 \right> & \left< v_2 v_2 \right> & \ldots & \left< v_2 v_n \right> \\ \vdots & \vdots & \ddots & \vdots \\ \left< v_n v_1 \right> & \left< v_n v_2 \right> & \ldots & \left< v_n v_n \right> \end{array} \right] - \bm \mu \bm \mu^\top \right).
\end{equation}
In every considered case each $v_i$ ($i \in \{1,2,...,n\}$) can be written as a linear combination of elements of the form $ n_j^{k_1} n_l^{k_2}$ where $j,l \in \{1,2,...,M_\textrm{pix} \}$,  and $k_1,k_2$ are natural exponents (possibly zero). Therefore, it's enough to be able to compute expected values of products $ \left< n_{j_1}^{k_1} n_{j_2}^{k_2} ... n_{j_r}^{k_r}  \right>$ for $r \le 4$ to reconstruct all terms of $\bm \mu$ and $\bm \Sigma$. The procedure used to compute these expected values is as follows. PSFs of both sources are numerically integrated over different pixels---we construct variables \begin{equation} U_{j,1} = \int_{x_j-\Delta x/2}^{x_j+\Delta x/2} U(x + \theta/2) dx ,~ U_{j,2} = \int_{x_j-\Delta x/2}^{x_j+\Delta x/2} U(x - \theta/2) dx, \end{equation} where $j\in \{ 1,2,...,M_\textrm{pix}\}$ denotes the pixel label. Now we use the fact, that the number of photons detected in each pixel, when the sources brightness are fixed, is described by a Poisson distribution with  a mean value
\begin{equation}
\label{mean_pois}
\left< n_j | q_1, q_2 \right>= \left( q_1 U_{j,1} + q_2 U_{j,2} \right) \bar{n},
\end{equation}
where $q_1, q_2 \in \{q_\textrm{off},q_\textrm{on} \}$ denote relative brightness of the 1st and the 2nd emitter respectively. When external,  Poissonian noise (studied in Section \ref{secnoise}) is present, the above equation should be modified to
\begin{equation}
\label{mean_pois_2}
\left< n_j | q_1, q_2 \right>= \left( q_1 U_{j,1} + q_2 U_{j,2} \right) \bar{n} + \mu_B,
\end{equation}
where $\mu_B$ is the mean value of this noise in each pixel in each frame.
Further steps remain valid, as conditional random variables $n_j|q_1,q_2$ are Poissonian and mutually independent with and without external noise. The expected value of a product of their powers is

\begin{equation}
\label{S42}
\left< n_{j_1}^{k_1} n_{j_2}^{k_2} ...n_{j_r}^{k_r}|q_1,q_2 \right> = \prod_{i=1}^r M_{k_i}\left(\left< n_{j_i} | q_1, q_2 \right> \right),
\end{equation}where $M_k(v)$ denotes a $k$-th  raw moment of Poisson distribution with a mean value $v$, e.g. $M_0(\nu) = 1$, $M_1(\nu)=\nu$, $M_2(\nu) = \nu^2 +\nu$, $M_3(\nu) = \nu^3+3 \nu^2+\nu$, $M_4(v)=\nu^4+6 \nu^3 + 7 \nu^2 + \nu$. Equation \eqref{S42} is only valid if indices $j_1,...,j_r$ are mutually different---if any index repeats, one should replace an expression of the form $n_j^{k_1} n_j^{k_2}$ with an expression $n_j^{k_1+k_2}$, repeat such a procedure as long as there are any repetitions left, and only then use \eqref{S42} directly. Already described steps allow us to compute conditional expected values. In order to compute the desired expected values $ \left< n_{j_1}^{k_1} n_{j_2}^{k_2} ... n_{j_r}^{k_r} \right>$ one only needs to average the conditional ones over four different configurations of the emitters using the formula
\begin{equation}
\left<X\right> = p^2 \left<X|q_\textrm{off}, q_\textrm{off} \right>+p(1-p) \left( \left<X|q_\textrm{off}, q_\textrm{on} \right>+\left<X|q_\textrm{on}, q_\textrm{off} \right> \right) +(1-p)^2 \left<X|q_\textrm{off}, q_\textrm{off} \right>.
\end{equation}\newline
By performing the described steps numerically, we obtain $\mathcal F(\theta)$ (and consequently $\zeta$) associated with different image reconstruction algorithms in the simplified blinking model.

We will now show, how to extend this scheme to the case of Markov process based realistic model. Let's first remind, that the relative brightness of each emitter is described by a Markov process with two possible states with different relative brightness: $q_\textrm{on}$ and $q_\textrm{off}$. The brightness of each emitter is a function of time $P_i(t)$ ($ i \in \{ 1,2\}$), which takes two possible values: $q_\textrm{off} \bar P$, $q_\textrm{on} \bar P$. During a short time interval $\left[t, t+\delta t\right]$ $i$-th emitter emits $ P_i(t) \delta t$ photons on average. Lifetimes of on- and off- states are equal to $\tau_\textrm{off}$ and $\tau_\textrm{on}$ respectively. The probability that a given emitter remains in a fixed state with a lifetime $\tau_i$ for a time period $t$ is proportional to $\exp(-t/\tau_i)$. Let's now introduce a more formal description of the Markov process, which leads to such an exponential behaviour. At any time $t$, the state of an emitter is described by a vector $\begin{bmatrix} p_\textrm{off} \\ p_\textrm{on} \end{bmatrix}$ , where $p_\textrm{off}$ and $p_\textrm{on}$ denote the probabilities of finding the emitter in off- and on- state respectively. The time evolution of the emitter state is given by
\begin{equation}
\begin{bmatrix} p_\textrm{off} \\ p_\textrm{on} \end{bmatrix}(t + \Delta t) = \bm T (\Delta t) \begin{bmatrix} p_\textrm{off} \\ p_\textrm{on} \end{bmatrix}(t),
\end{equation}
where $ \bm T (\Delta t)$ is a transition matrix defined as
\begin{equation}
\bm T(\Delta t) = \begin{bmatrix}t_{00} & t_{10} \\ t_{01} & t_{11} \end{bmatrix}(\Delta t) =\exp \left( \Delta t \begin{bmatrix} -\tau_\textrm{off}^{-1} & \tau_\textrm{on}^{-1} \\ \tau_\textrm{off}^{-1} & -\tau_\textrm{on}^{-1} \end{bmatrix}  \right) .
\end{equation}
It's easy to check, that after a long evolution the state always converges to
\begin{equation}
\begin{bmatrix} p_\textrm{off} \\ p_\textrm{on} \end{bmatrix}(\Delta t \rightarrow \infty)= \begin{bmatrix} \tilde p_\textrm{off} \\ \tilde p_\textrm{on} \end{bmatrix} = \frac{1}{\tau_\textrm{off} + \tau_\textrm{on}}\begin{bmatrix} \tau_\textrm{off} \\ \tau_\textrm{on} \end{bmatrix}.
\end{equation}
$\begin{bmatrix} \tilde p_\textrm{off} \\ \tilde p_\textrm{on} \end{bmatrix}$ is a stationary state of the process, and will be used as an initial state in our considerations---if no information about the previous run of the process is available, the probability of finding an emitter in a given state is proportional to its lifetime. Using the transition matrix $\bm T(\Delta t)$ elements, let's define the $\bm S(\Delta t)$ matrix:
\begin{equation}
\bm S(\Delta t) = \begin{bmatrix}t_{00}(\Delta t)  q_\textrm{off} \bar P & t_{01}(\Delta t)  q_\textrm{off} \bar P \\ t_{10}(\Delta t)  q_\textrm{on} \bar P & t_{11}(\Delta t)  q_\textrm{on} \bar P \end{bmatrix},
\end{equation}which allows us to write down formulas for temporal brightness correlations in a compact way:
\begin{equation}
\label{S48}
\left< P_i(t_1) P_i(t_2) ... P_i(t_r) \right> = \begin{bmatrix} \tilde p_\textrm{off} & \tilde p_\textrm{on} \end{bmatrix} \bm{S}(t_2-t_1) \bm{S}(t_3-t_2)... \bm{S}(t_r-t_{r-1})  \begin{bmatrix} q_\textrm{off} \bar P \\ q_\textrm{on} \bar P \end{bmatrix}.
\end{equation}
In the above formula $t_1 \le t_2 \le ... \le t_r$, and $\left< \bullet \right>$ denotes averaging over Markov processes $P_1(t)$, $P_2(t)$. Two emitters are independent, so in order to compute a product in which $P_1$ and $P_2$ terms are mixed, one can use the formula
\begin{equation}
\left< \mathcal G_1 \left [P_1(t) \right] \mathcal G_2 \left [P_2(t) \right] \right> = \left< \mathcal G_1 \left [P_1(t) \right] \right> \left< \mathcal G_2 \left [P_2(t) \right] \right>,
\end{equation}
which is true for all functionals $\mathcal G_1$, $\mathcal G_2$. Further on, the following integrals of correlations over detection time frames will be useful:
\begin{equation}
\label{S50}
\chi_1 = \int_0^\tau \left<P_i(t)\right> \textrm{d}t = \left<P_i\right> \tau
\end{equation}
\begin{equation}
\chi_{2,m} = \int_0^\tau \textrm{d}t_1 \int_{(m-1) \tau}^{m \tau} \textrm{d}t_2 \left< P_i(t_1) P_i(t_2) \right>
\end{equation}
\begin{equation}
\chi_{3,m} = \int_0^\tau \textrm{d}t_1 \int_0^\tau dt_2 \int_{(m-1) \tau}^{m \tau} \textrm{d}t_3 \left< P_i(t_1) P_i(t_2) P_i(t_3) \right>
\end{equation}
\begin{equation}
\label{S53}
\chi_{4,m} = \int_0^\tau \textrm{d}t_1 \int_0^\tau \textrm{d}t_2 \int_{(m-1) \tau}^{m \tau} \textrm{d}t_3 \int_{(m-1) \tau}^{m \tau} \textrm{d}t_4 \left< P_i(t_1) P_i(t_2) P_i(t_3) P_i(t_4) \right>
\end{equation}
At this point, we are prepared to attack the problem of computing $\mathcal F$. Although the frames are now correlated, we still consider vectors $\bm N$ that can be written as in \eqref{6}, and we can use the central limit theorem in its extended version \cite{Rosenblatt43} because correlations between frames decay exponentially with time. Therefore, it's again enough to calculate $\bm \mu$ and $\bm \Sigma$ associated with $\bm N$, and then use \eqref{S36}. The scaling of both terms in this equation remains the same, so the 2nd term again disappears in the limit $M_\textrm{fr} \rightarrow \infty$. \eqref{mu} is still valid, and can be used to calculate $\bm \mu$, but in order to compute $ \bm \Sigma$ elements one needs to take into account correlations between frames:
\begin{equation}
\bm \Sigma_{ij} = \frac{1}{M_\textrm{fr}^2} \sum_{m,m'=1}^{M_\textrm{fr}} \textrm{cov}(v_{i,m}, v_{j,m'}).
\end{equation}
In the limit $M_\textrm{fr} \rightarrow \infty$, using the homogeneity of the Markov processes, we can simplify our formula:
\begin{equation}
\label{S55}
\bm \Sigma_{ij} = \frac{1}{M_\textrm{fr}} \left(  \textrm{cov}(v_{i,1},v_{j,1})+2 \sum_{m=2}^{\infty} \textrm{cov}(v_{i,1}, v_{j,m}) \right).
\end{equation}
Analogously to the previous case, photon numbers in different pixels and time frames are uncorrelated and described by a Poisson distribution if functions $P_1(t), P_2(t)$ are fixed, and we have:
\begin{equation}
\left< n_{j,m} |P_1(t), P_2(t) \right> \equiv \mu_{j,m} = \int_{(m-1)\tau}^{m \tau} \left( P_1(t) U_{1,j}  + P_2(t) U_{2,j}\right) dt.
\end{equation}Conditional products of variables $n_{j,m}$ are computed with the help of the formula
\begin{equation}
\label{S57}
\left< n_{j_1, m_1}^{k_1} n_{j_2, m_2}^{k_2} ...n_{j_r, m_r}^{k_r}|P_1(t), P_2(t) \right> = \prod_{i=1}^r M_{k_i}\left(\mu_{j_i,m_i} \right)
\end{equation}
valid if pairs $(j_i, m_i)$ mutually differ on at least one position. All terms of $\bm \mu$ and $\bm \Sigma$ are linear combinations of expectation values (where averaging over Markov processes is made) $\left< n_{j_1, m_1}^{k_1} n_{j_2, m_2}^{k_2} ...n_{j_r, m_r}^{k_r} \right> $. We want restrict ourselves to $ \bm N $ which consist of 2nd order correlations at most, so from now on we assume that $k_1+k_2+...k_r \le 4$. Then, after expanding the RHS of \eqref{S57}, and using formulas for Poisson distribution moments, we see that all conditional expected values are linear combinations of products of $\mu_{j,m}$ with at most 4 terms. Expectation values required to reconstruct $\bm \mu$ and $\Sigma$ are linear combinations of similar products averaged over Markov processes. Such products can be written with the help of variables $\chi$, defined  in \eqref{S50}-(\ref{S53}), 
\begin{equation}
\label{S58}
 \left<\mu_{j,m} \right> = (U_{1,j} + U_{2,j}) \chi_1,
 \end{equation}
\begin{equation}
\left< \mu_{j,0} \mu_{j',m} \right> = (U_{1,j}U_{1,j'} + U_{2,j}U_{2,j'}) \chi_{2,m} + (U_{1,j} U_{2,j'} + U_{2,j} U_{1,j'}) \chi_1^2,
\end{equation}
\begin{multline}
\left< \mu_{j,0} \mu_{j',0} \mu_{j'',m} \right> = U_{1,j}U_{1,j'}U_{ 1,j''} \chi_{3,m} +U_{1,j}U_{1,j'}U_{ 2,j''} \chi_{2,0}\chi_1 + \\ +(U_{1,j}U_{2,j'}U_{ 1,j''}+U_{1,j}U_{2,j'}U_{ 2,j''}) \chi_{2,m} \chi_1 + (1 \leftrightarrow 2),
\end{multline}
\begin{multline}
\label{S61}
\left< \mu_{j,0} \mu_{j',0} \mu_{j'',m} \mu_{j''',m} \right> = U_{1,j}U_{1,j'}U_{ 1,j''}U_{ 1,j'''} \chi_{4,m} +\\ +(U_{1,j}U_{1,j'}U_{ 1,j''}U_{ 2,j'''}+U_{1,j}U_{1,j'}U_{ 2,j''}U_{ 1,j'''}+U_{1,j}U_{2,j'}U_{ 1,j''}U_{ 1,j'''}+U_{1,j}U_{2,j'}U_{ 2,j''}U_{ 2,j'''})\chi_{3,m} \chi_1 + \\ + U_{1,j}U_{1,j'}U_{ 2,j''}U_{ 2,j'''} \chi_{2,0}^2 + (U_{1,j}U_{2,j'}U_{ 1,j''}U_{ 2,j'''}+U_{1,j}U_{2,j'}U_{ 2,j''}U_{ 1,j'''}) \chi_{2,m}^2 +(1 \leftrightarrow 2).
\end{multline}
Notation $+(1 \leftrightarrow 2)$ means that terms with swapped indices $1$ and $2$, that correspond to the 1st and 2nd emitter, should be added. 

Let's summarize the procedure used to compute $\bm \mu$ and $\bm \Sigma$ for the Markov process based model. First, (\ref{mu}) and (\ref{S55}) are applied, and all terms are expressed as linear combinations of expected values $\left< n_{j_1, m_1}^{k_1} n_{j_2, m_2}^{k_2} ...n_{j_r, m_r}^{k_r} \right> $. Then, conditional expected values $\left< n_{j_1, m_1}^{k_1} n_{j_2, m_2}^{k_2} ...n_{j_r, m_r}^{k_r}|P_1(t), P_2(t) \right> $ are computed with the help of \eqref{S57}. Averaging over Markov processes $P_1(t)$, $P_2(t)$ is made after expanding the RHS of \eqref{S57}, formulas (\eqref{S58}-\eqref{S61}) are then utilized to express $\left< n_{j_1, m_1}^{k_1} n_{j_2, m_2}^{k_2} ...n_{j_r, m_r}^{k_r} \right> $ using variables $U_{1,j}$, $U_{2,j}$, and $\chi$ (defined in \eqref{S50}-(\ref{S53})). Notice, that the sum present in \eqref{S58} is infinite, but in our case all sums converge. Moreover, it's possible to express all the terms of $\bm \mu$ and $ \bm \Sigma$ in the considered cases with the help of variables $U_{1,j}, U_{2,j}$ (computed numerically), $\chi_1$, $\chi_{2,1}$, $\chi_{3,1}$, $\chi_{4,1}$, and the following infinite sums (all of them converge):
\begin{equation}
S_1 = \sum_{m=2}^\infty\left( \chi_{2,m} - \chi_1^2 \right),
\end{equation}
\begin{equation}
S_2 = \sum_{m=2}^\infty\left( \chi_{3,m} - \chi_{2,1} \chi_1 \right),
\end{equation}
\begin{equation}
S_3 = \sum_{m=2}^\infty\left( \chi_{4,m} - \chi_{2,1}^2 \right),
\end{equation}
\begin{equation}
S_4 = \sum_{m=2}^\infty\left( \chi_{2,m}^2 - \chi_1^4 \right)
\end{equation}
where $\chi$ variables are expressed as functions of the setup parameters analytically with the help of (\ref{S48}),(\ref{S50})-(\ref{S53}), and then sums $S_1, S_2, S_3, S_4$ are also computed analytically (they can be expressed as sums of geometric series).

\subsection{Proof of \eqref{S32}}
\label{secD}
In this section we provide a rigorous proof of \eqref{S32}. For $\alpha=0$, we have $A=0$, the value of the sum doesn't depend on $\bar{n}$ and can be easily computed, and the proof becomes trivial. For $0<\alpha \le 1$ we need to prove that
\begin{equation}
    \lim_{\bar{n} \rightarrow \infty} \sum_{n=0}^{\infty} \frac{e^{-\bar{n}} \bar{n} ^n}{n!} \left( B e^{A \bar{n}}(1-A)^n+C+D e^{-A \bar{n}} (1+A)^n\right)^{-1} = C^{-1},
\end{equation}
where $0<A\le 1$, $B,D \ge 0$, $C>0$. Let us define
\begin{equation}
X_n(\bar{n}) := \frac{e^{-\bar{n}} \bar{n} ^n}{n!} \left( B e^{A \bar{n}}(1-A)^n+C+D e^{-A \bar{n}} (1+A)^n\right)^{-1} .
\end{equation}
 We are going to use the following property of Poisson distribution:
\begin{equation}
\label{pois}
\forall _{ \delta >0, \epsilon >0} ~ \exists_{N_0} ~ \forall_{\bar{n} > N_0} : ~ 0 <  \sum_{n=0}^\infty \frac{ e^{- \bar{n}} \bar{n} ^n}{n!} -\sum_{n=(1-\epsilon) \bar{n}}^{(1+\epsilon) \bar{n}} \frac{ e^{- \bar{n}} \bar{n} ^n}{n!} < \delta.
\end{equation}
Intuitively, for large enough $\bar{n}$, all probable values of Poisson distribution are located in the range $ \bar{n} (1 \pm \epsilon)$, because the standard deviation of Poisson distribution with mean $\bar{n}$ is $\sqrt{\bar{n}}$ . Let's notice that $X_n(\bar{n} ) \le C^{-1} \frac{\bar{n}^n e^{- \bar{n}}}{n!}$, so the following inequality is true:
\begin{equation}
0<\sum_{n=0}^\infty X_n(\bar{n}) -\sum_{n=(1-\epsilon) \bar{n}}^{(1+\epsilon) \bar{n}} X_n( \bar{n})  \le C^{-1} \left( \sum_{n=0}^\infty \frac{ e^{- \bar{n}} \bar{n} ^n}{n!} -\sum_{n=(1-\epsilon) \bar{n}}^{(1+\epsilon) \bar{n}} \frac{ e^{- \bar{n}} \bar{n} ^n}{n!}   \right),
\end{equation}which after using \eqref{pois} allows us to conclude that:
\begin{equation}
\forall _{  \epsilon >0} ~ \lim_{\bar{n} \rightarrow \infty} \sum_{n=0}^\infty X_n(\bar{n}) = \lim_{\bar{n} \rightarrow \infty} \sum_{n=(1-\epsilon) \bar{n}}^{(1+\epsilon) \bar{n}} X_n(\bar{n}).
\end{equation}
Let's now fix $\epsilon , \delta_1 > 0$ satisfying
\begin{equation}
\label{in1}
A+(1-\epsilon) \log (1-A) <- \delta_1,
\end{equation}
\begin{equation}
\label{in2}
-A+(1+\epsilon) \log (1+A) < -\delta_1.
\end{equation}
For $A=1$ the first inequality may be omitted. It's easy to show, that the described choice of positive constants $\epsilon$ and $\delta_1$  is always possible. Such a choice allows us to conclude, that for $n \in \left[ \bar{n} (1-\epsilon), \bar{n} (1+\epsilon) \right]$:
\begin{equation}
e^{A \bar{n}} (1-A)^n \le e^{\bar{n} (A+(1-\epsilon) \log (1-A))} < e^{- \delta_1 \bar{n}},
\end{equation}
\begin{equation}
e^{-A \bar{n}} (1+A)^n \le e^{\bar{n} (-A+(1+\epsilon) \log (1+A))}< e^{- \delta_1 \bar{n}}.
\end{equation}
Therefore,
\begin{equation}
1 \le \left( \frac{ e^{- \bar{n}} \bar{n} ^n}{ C n!} \right) / X_n(\bar{n}) \le 1 + \frac{B+D}{C} e^{- \delta_1 \bar{n}},
\end{equation}
and consequently
\begin{equation}
\left( 1+\frac{B+D}{C} e^{-\delta_1 \bar{n}} \right)^{-1} \sum_{n=(1-\epsilon) \bar{n}}^{(1+\epsilon) \bar{n}} \frac{ e^{- \bar{n}} \bar{n} ^n}{ C n!} \le \sum_{n=(1-\epsilon) \bar{n}}^{(1+\epsilon) \bar{n}} X_n(\bar{n}) \le \sum_{n=(1-\epsilon) \bar{n}}^{(1+\epsilon) \bar{n}} \frac{ e^{- \bar{n}} \bar{n} ^n}{ C n!}.
\end{equation}
Since $e^{-\delta_1 \bar{n}} \rightarrow 0$ for $\bar{n} \rightarrow \infty$, and sum $\sum_{n=(1-\epsilon) \bar{n}}^{(1+\epsilon) \bar{n}} \frac{ e^{- \bar{n}} \bar{n} ^n}{ C n!}$ converges to $C^{-1}$ (as a consequence of \eqref{pois}), we can use the so called \textit{sandwich theorem} to finally conclude that
\begin{equation}
\lim_{\bar{n} \rightarrow \infty} \sum_{n=0}^\infty X_n(\bar{n}) = \lim_{\bar{n} \rightarrow \infty} \sum_{n=(1-\epsilon) \bar{n}}^{(1+\epsilon) \bar{n}} X_n(\bar{n})=C^{-1}.
\end{equation}

\section*{References}

\bibliographystyle{unsrt}
\bibliography{sofi_estimation}

\end{document}